\documentclass[journal,twocolumn]{IEEEtran}

\usepackage{acro}
\acsetup{make-links=true,single=true}

\usepackage{amsmath,amssymb,amsfonts}
\usepackage{bm}
\usepackage{graphicx}
\usepackage{cite}
\usepackage[vlined]{algorithm2e}
\usepackage{tikz}
\usetikzlibrary{arrows.meta,positioning,shapes.geometric,calc,patterns}
\usepackage{hyperref}
\usepackage{pgfplots}
\usepackage{comment}
\pgfplotsset{compat=1.18}

\RestyleAlgo{boxruled}
\SetAlCapFnt{\small}
\SetAlCapNameFnt{\small\bfseries}
\SetAlgoCaptionSeparator{.}
\SetAlgoNlRelativeSize{-1}

\SetNlSty{textbf}{\hspace*{0.75em}\scriptsize}{\hspace*{0.75em}}
\SetNlSkip{1.0em}

\SetKwInput{KwIn}{Input}
\SetKwInput{KwOut}{Output}
\SetInd{0.50em}{0.90em}
\DontPrintSemicolon

\DeclareAcronym{ABLS}{
  short=ABLS,
  long=alternating best-improvement local search
}

\DeclareAcronym{AWGN}{short=AWGN, long=additive white Gaussian noise}
\DeclareAcronym{AIR}{short=AIR, long=achievable information rate}
\DeclareAcronym{AO}{short=AO, long=alternating optimization}

\DeclareAcronym{CSI}{short=CSI,long=channel state information}
\DeclareAcronym{DL}{short=DL, long=downlink}
\DeclareAcronym{DODS}{short=DODS, long=direct one-/two-bit discrete search}

\DeclareAcronym{EE}{short=EE, long=energy efficiency}

\DeclareAcronym{FDPASS}{short=FD-PASS, long=full-duplex pinching-antenna system}
\DeclareAcronym{FAS}{short=FAS, long=fluid antenna system}
\DeclareAcronym{HD}{short=HD, long=half-duplex}
\DeclareAcronym{IBFD}{short=IBFD, long=in-band full-duplex}
\DeclareAcronym{iid}{
  short=i.i.d.,
  long=independent and identically distributed
}
\DeclareAcronym{ISAC}{short=ISAC, long=integrated sensing and communication}

\DeclareAcronym{LoS}{short=LoS, long=line-of-sight}

\DeclareAcronym{MB}{short=MB, long=Maxwell--Boltzmann}
\DeclareAcronym{MI}{short=MI, long=mutual information}
\DeclareAcronym{MIMO}{short=MIMO, long=multiple-input multiple-output}
\DeclareAcronym{ML}{short=ML, long=maximum likelihood}
\DeclareAcronym{MM}{short=MM, long=majorization--minimization}
\DeclareAcronym{MMSE}{short=MMSE, long=minimum mean square error}
\DeclareAcronym{MISOCP}{
  short = MISOCP,
  long  = mixed-integer second-order-cone program
}
\DeclareAcronym{NOMA}{short = NOMA,long = non-orthogonal multiple access}
\DeclareAcronym{PE}{short=PE, long=pinching element}
\DeclareAcronym{PAPR}{short=PAPR, long=peak-to-average power ratio}
\DeclareAcronym{PASS}{short=PASS, long=pinching-antenna system}
\DeclareAcronym{PDF}{short=PDF, long=probability density function}
\DeclareAcronym{PMF}{short=PMF, long=probability mass function}
\DeclareAcronym{PS}{short=PS, long=probabilistic shaping}

\DeclareAcronym{QAM}{short=QAM, long=quadrature amplitude modulation}
\DeclareAcronym{QPSK}{short=QPSK, long=quadrature phase-shift keying}

\DeclareAcronym{RF}{
  short=RF,
  long=radio frequency
}
\DeclareAcronym{RX}{short=RX, long=receiver}

\DeclareAcronym{SCA}{short=SCA, long=successive convex approximation}
\DeclareAcronym{SE}{short=SE, long=spectral efficiency}
\DeclareAcronym{SI}{short=SI, long=self-interference}
\DeclareAcronym{SNR}{short=SNR, long=signal-to-noise ratio}
\DeclareAcronym{SS}{short=SS, long=super symbol}
\DeclareAcronym{STAR}{short=STAR, long=simultaneous transmit and receive}
\DeclareAcronym{SOCP}{
  short=SOCP,
  long=second-order cone program
}

\DeclareAcronym{TX}{short=TX, long=transmitter}
\DeclareAcronym{TDD}{
  short=TDD,
  long=time-division duplexing
}
\DeclareAcronym{UL}{short=UL, long=uplink}

\title{FullPASS: Geometry Optimization for Full-Duplex Pinching-Antenna Systems}

\author{Morteza Barzegar Astanjin,~\IEEEmembership{Student Member,~IEEE},
Seyed Mohammad Azimi-Abarghouyi,~\IEEEmembership{Member,~IEEE},
Thomas Eriksson,~\IEEEmembership{Senior Member,~IEEE},
and Mikko Valkama,~\IEEEmembership{Fellow,~IEEE}%
\thanks{Morteza Barzegar Astanjin, Seyed Mohammad Azimi-Abarghouyi, and Thomas Eriksson are with the Department of Electrical Engineering, Chalmers University of Technology, Gothenburg, Sweden (e-mail: {mortezab, azimimo, thomase}@chalmers.se).}%
\thanks{Mikko Valkama is with the Faculty of Information Technology and Communication Sciences, Tampere University, Tampere, Finland (e-mail: mikko.valkama@tuni.fi).}%
}

\begin{document}

\maketitle

\begin{abstract}
    This paper proposes FullPASS, an in-band full-duplex architecture for pinching-antenna systems (PASSs) based on two parallel waveguides. The FullPASS transceiver simultaneously communicates with a single full-duplex user terminal over the same time--frequency resource: the transmit waveguide delivers the downlink signal, while the receive waveguide collects the uplink signal. Candidate pinching elements are placed along both waveguides, and the FullPASS transceiver jointly selects the active transmit and receive elements. We derive a geometry-based channel model for the downlink, uplink, and transmit-to-receive self-interference paths, including free-space propagation, in-waveguide propagation phase and attenuation, and the attenuation caused by upstream activated elements along each waveguide. The joint activation problem is formulated as a binary optimization that maximizes the bidirectional sum spectral efficiency between FullPASS and the full-duplex user terminal while keeping the self-interference leakage at the FullPASS receiver below a prescribed threshold. The coupled binary activation decisions and activation-dependent waveguide
attenuation make the selection problem nonconvex and combinatorial. To solve it, we develop a two-stage algorithm. The first stage uses phase-anchored second-order cone relaxations and deterministic rounding to generate binary trial activation patterns over a simplified propagation model. The second stage applies alternating best-improvement local search with add, remove, and swap operations evaluated under the full propagation model. We also characterize the dominant computational cost, showing that the proposed method replaces exhaustive enumeration with a finite number of conic subproblems and local candidate evaluations. Simulations show that the proposed method achieves an average sum spectral
efficiency within $0.95\%$ of exhaustive search on both the $13\times13$ and
$15\times15$ candidate grids. At the $15\times15$ grid, it reduces the average
runtime by more than one order of magnitude relative to exhaustive search, and
it remains applicable to substantially larger candidate sets where exhaustive
enumeration is not evaluated.
\end{abstract}


\section{Introduction}

There is growing interest in flexible and reconfigurable antenna architectures due to their potential to dynamically adapt wireless systems to user mobility, signal blockages, and complex propagation environments. Conventional fixed antenna arrays provide spatial processing gain, but their radiating locations are determined at deployment and cannot be easily adapted to changes in the environment. This limitation is especially relevant at high carrier frequencies, where path loss, blockage, and geometric alignment strongly influence the link quality. Recent concepts such as fluid-antenna systems, movable-antenna systems, and reconfigurable intelligent surfaces address this issue by introducing different forms of spatial or propagation reconfigurability \cite{new2025fas,ning2024movable,samy2025passris}. 

Building on these principles of spatial adaptability, \ac{PASS} has recently emerged as a distinct, waveguide-based approach to spatial reconfiguration. First proposed and prototyped by NTT DOCOMO in 2021 \cite{fukuda2022pinching}, \ac{PASS} is considerably newer than the reconfigurable technologies discussed above, and it has attracted rapidly growing research interest only over the past two years \cite{ding2025pinchingperspective,yang2025principles,liu2025passarchitecture,liu2025passtutorial,wang2025modeling,ouyang2026capacity,zhao2026wdma}. It also reconfigures the antenna geometry in a fundamentally different way: fluid- and movable-antenna systems relocate radiating elements within a compact aperture spanning only a few wavelengths, and a reconfigurable intelligent surface reshapes propagation passively while leaving the source fixed and incurring the compounded loss of a cascaded link, whereas a \ac{PASS} transports the signal along a low-cost dielectric waveguide and couples it to free space at selected \acp{PE} that can be dynamically activated or physically relocated over a waveguide extent spanning meters. By creating radiating or receiving locations on demand close to the user, a \ac{PASS} establishes short line-of-sight links, mitigates large-scale path loss and blockage, and flexibly scales the number and position of active antennas at low cost and without dedicated phase-shifting hardware. These properties allow a \ac{PASS} to efficiently support user-centric coverage, beamforming, and multiple-access operation \cite{ding2025pinchingperspective,yang2025principles,liu2025passarchitecture,liu2025passtutorial,wang2025modeling,ouyang2026capacity,zhao2026wdma}.

Previous studies on \ac{PASS} have extensively explored optimizing continuous \ac{PE} positions, beamforming, user association, robust design, secure communications, and channel estimation \cite{xu2025downlinkrate,tegos2025uplinkmin,zhang2025twotimescale,zeng2025eeuplink,zeng2025robust,jiang2025covert,xiao2025channel,guo2025gpass}. A complementary line of work considers discrete activation and finite-candidate \ac{PE} selection, including activation design and \ac{NOMA}-assisted schemes \cite{wang2025nomaactivation,wang2025activation,cao2026discretepass,zhao2026wdma,khisa2025joint}. Such discrete formulations are particularly relevant for practical implementations, where feasible \ac{PE} locations are limited by placement resolution, mechanical activation mechanisms, minimum-spacing requirements, and hardware constraints. These selection problems become substantially harder once transmit and receive \acp{PE} must be chosen jointly under simultaneous same-band operation, since the same binary activation decisions then determine the desired uplink and downlink channels and the transmit-to-receive \ac{SI} coupling at once. Efficiently solving this coupled binary selection problem is the central focus of this paper.

This \ac{SI} challenge is precisely the one addressed by \ac{IBFD} communication, a proven framework for improving spectral efficiency by enabling simultaneous transmission and reception on the same time--frequency resource. Experimental and theoretical studies have established the feasibility of \ac{IBFD} operation through a combination of antenna-domain isolation and transceiver-side \ac{SI} mitigation techniques \cite{choi2010single,jain2011practical,bharadia2013full,sabharwal2014inband,zhang2016fdchallenges}. Nevertheless, \ac{SI} remains the main challenge because the transmitted signal can be many orders of magnitude stronger than the desired received signal. In multi-antenna, wireless local-area-network, massive \ac{MIMO}, and distributed \ac{MIMO} settings, the attainable full-duplex gain depends not only on the overall suppression level but also on the spatial coupling between transmit and receive radio chains \cite{aryafar2012midu,bharadia2014mimo,duarte2014design,riihonen2011loopback,sultan2021jointfd,mohammadi2024ten}. 

Recognizing that this spatial coupling is a critical degree of freedom, full-duplex architectures employing reconfigurable antennas have recently begun to attract attention. In particular, fluid-antenna-based full-duplex studies have exploited antenna-position flexibility to influence the \ac{SI} channel, while recent full-duplex \ac{FAS}-assisted designs jointly optimize transmit and receive antenna positions together with transceiver variables \cite{skouroumounis2023fluidfd,hong2025fassic,tang2026fdfasisac}. These works demonstrate that spatial reconfigurability can provide an additional degree of freedom for \ac{SI} management, but they confine that freedom to small apertures of a few wavelengths, which limits how much transmit--receive isolation geometry alone can provide. \ac{PASS} is particularly well suited to overcome this limitation: because a waveguide transports the signal and the active \ac{PE} determine where power is coupled from the guided wave into free space, the transmit and receive radiating elements can be chosen over a waveguide extent spanning meters rather than repositioned within a local aperture. This large geometric separation lets a full-duplex \ac{PASS} transceiver place transmit and receive elements at weakly coupled locations, providing passive propagation-domain self-interference suppression before analog or digital cancellation, while using only low-cost \acp{PE} instead of dedicated isolation hardware. At the same time, the selected transmit and receive locations shape the desired downlink and uplink channels, so the same discrete \ac{PE} selection simultaneously governs desired-signal quality and \ac{SI} leakage. This coupling is precisely what makes the selection problem both powerful and nontrivial, and it motivates studying \ac{PE} selection as a propagation-domain \ac{SI}-suppression mechanism that complements conventional full-duplex isolation and cancellation techniques.

Integrating full-duplex capability fundamentally changes the \ac{PASS} design problem relative to existing half-duplex formulations. In the proposed FullPASS architecture, a two-waveguide transceiver exchanges downlink and uplink signals with a single full-duplex user terminal over the same time--frequency resource, and it must decide which transmit and receive \acp{PE} to activate. We restrict attention to propagation-domain \ac{SI} suppression at the FullPASS transceiver, treating terminal-side \ac{SI} leakage as independently controlled and unaffected by the selected \acp{PE}. The core difficulty is that one set of binary activation decisions simultaneously governs three coupled effects: the strength of the desired downlink and uplink channels; the guided-wave attenuation along each waveguide, comprising a fixed per-meter propagation loss and an activation-dependent pass-through loss from upstream active elements; and the transmit-to-receive \ac{SI} between the two waveguides. As a result, the natural design goal—maximizing the desired bidirectional spectral efficiency while keeping \ac{SI} leakage below a prescribed threshold—yields a highly nonconvex, combinatorial selection problem in which desired-signal quality and \ac{SI} cannot be tuned independently, and whose exhaustive solution scales poorly with the number of candidate locations.

These challenges—the coupled binary decisions, the practical waveguide attenuation, and the \ac{SI} constraint—call for a design that treats the FullPASS architecture, its propagation model, and the resulting element-selection problem together. Accordingly, the main contributions of this paper are summarized as follows:

\begin{itemize}
\item \emph{Full-duplex \acs{PASS} architecture with propagation-domain passive
\acs{SI} suppression through pinching-element selection.}
We introduce FullPASS, an in-band full-duplex \ac{PASS} architecture for
point-to-point bidirectional communication with a single full-duplex user
terminal over the same time--frequency resource, realized through a transmit
and a receive waveguide with independently selectable \acp{PE}. Its
defining feature is that transmit--receive isolation is treated as a
geometry-based pinching-element selection problem: the choice of active transmit and receive
elements shapes the desired downlink and uplink channels while directly
regulating the coupled \ac{SI}. This passive \ac{SI}-suppression mechanism complements conventional analog and digital self-interference cancellation.

\item \emph{Practical propagation model and selection problem.} We derive a
geometry-based line-of-sight channel model that jointly characterizes the
downlink, uplink, and pairwise transmit-to-receive \ac{SI} coupling paths.
Unlike idealized lossless treatments, each complex coefficient captures
free-space spherical spreading and phase, in-waveguide phase shift, and
continuous per-meter in-waveguide attenuation, while an activation-dependent
pass-through attenuation accounts for the guided power extracted by upstream active \acp{PE}. Building on this model, we formulate joint transmit--receive element selection
as a binary optimization that maximizes the bidirectional sum spectral
efficiency subject to a constraint keeping the coupled \ac{SI} below a
prescribed threshold.

\item \emph{Two-stage selection algorithm.} We develop a two-stage method to
solve the resulting problem efficiently. Stage~1 operates on a simplified model and solves a \emph{phase-anchored} \ac{SOCP}
relaxation---projecting the complex desired-channel combination onto a finite
grid of reference phases to obtain convex subproblems---together with an
empirically calibrated guard margin and deterministic rounding that yield
binary trial patterns that are feasible under the practical model. Stage~2 applies an \emph{\ac{ABLS}}  under the full practical model,
evaluating add, remove, and swap moves that strictly preserve the \ac{SI}-leakage constraint. The same \ac{ABLS} procedure also serves as a
standalone baseline, isolating the gain contributed by the \ac{SOCP}
initialization.

\item \emph{Complexity analysis and evaluation.} We characterize the dominant
computational cost of the method, showing that it replaces exponential
exhaustive enumeration with a finite number of conic subproblems and low-order
local candidate evaluations. Extensive simulations show that the method attains
near-optimal desired-link spectral efficiency---within $0.95\%$ of exhaustive
search on the nominal grid, recovering roughly three-quarters of the optimality
gap left by standalone local search---and remains applicable to large candidate
grids where exhaustive search becomes computationally intractable. We further
report parameter sweeps over the \ac{SI}-suppression requirement, coupling
strength, candidate density, and in-waveguide attenuation.
\end{itemize}

The remainder of this paper is organized as follows. Section~II presents the
FullPASS system and propagation model. Section~III formulates the practical
selection problem. Section~IV develops the proposed two-stage selection framework,
including the simplified model, phase-anchored \ac{SOCP} initialization,
and \ac{ABLS} refinement, followed by a complexity analysis. Section~V provides numerical results, and
Section~VI concludes the paper.

\section{System Model}
\label{sec:system_model}

\subsection{Dual-Waveguide Architecture}

\begin{figure}[t!]
\centering
\resizebox{0.95\columnwidth}{0.18\textheight}{\begin{tikzpicture}[
> =Latex,
 font=\small\sffamily,
 wavecol/.style={color=teal!80!black},
 pa/.style={
 draw=gray!60, fill=gray!10, thick, rounded corners=0.5pt,
 minimum width=0.12cm, minimum height=0.32cm, inner sep=0pt
 },
activepa/.style={
 draw=red!80!black, fill=red!60, thick, rounded corners=0.5pt,
 minimum width=0.12cm, minimum height=0.32cm, inner sep=0pt
 },
 userFD/.style={
 circle, draw=violet!80!black, fill=violet!20, thick,
 minimum size=0.48cm, inner sep=0pt
 },
 wave/.style={line width=3.5pt, wavecol, line cap=round},
 dim/.style={gray!80!black, thick}
 ]

% =====================================================
% Coordinates
% =====================================================
\coordinate (TXleft) at (-2.4, 2.65);
\coordinate (TXright) at ( 2.8, 2.65);

\coordinate (RXleft) at (-2.6, 1.70);
\coordinate (RXright) at ( 2.6, 1.70);

\coordinate (AreaA) at (-4.3, 1.0);
\coordinate (AreaB) at ( 4.1, 1.0);
\coordinate (AreaC) at ( 3.5, -2.15);
\coordinate (AreaD) at (-4.9, -2.15);

% =====================================================
% Service Region
% =====================================================
\filldraw[fill=gray!10, draw=gray!40, thick, rounded corners=2pt]
(AreaA) -- (AreaB) -- (AreaC) -- (AreaD) -- cycle;
\node[font=\bfseries\small, text=gray!32, align=center]
at (-2.8, -1.7) {SERVICE REGION};

% =====================================================
% FullPASS Transceiver
% =====================================================
\node[
draw=gray!80!black, fill=gray!5, thick, rounded corners=3pt,
minimum width=1.1cm, minimum height=1.7cm,
align=center, font=\bfseries\scriptsize
] (TRX) at (-3.7, 2.175) {FullPASS\\TRX};

\draw[line width=1.5pt, wavecol] (TRX.east |- TXleft) -- (TXleft);
\draw[line width=1.5pt, wavecol] (TRX.east |- RXleft) -- (RXleft);

% =====================================================
% Waveguides and Candidate PAs
% =====================================================
\draw[wave] (TXleft) -- (TXright);
\node[above, font=\bfseries\scriptsize, wavecol] at (0, 2.85)
{TX Waveguide};

\draw[wave] (RXleft) -- (RXright);
\node[below, font=\bfseries\scriptsize, wavecol] at (0, 1.50)
{RX Waveguide};

\foreach \x in {-2.35, -2.10, -1.85, -1.60, -1.35, -1.10, -0.85,
-0.60, -0.35, -0.10, 0.15, 0.40, 0.65, 0.90, 1.15,
1.40, 1.65, 1.90, 2.15, 2.40}{
\node[pa] at (\x, 1.70) {};
}

\foreach \x in {-2.15, -1.90, -1.65, -1.40, -1.15, -0.90, -0.65,
-0.40, -0.15, 0.10, 0.35, 0.60, 0.85, 1.10, 1.35,
1.60, 1.85, 2.10, 2.35, 2.60}{
\node[pa] at (\x, 2.65) {};
}

% Activated PAs
\node[activepa] (tx1) at (-1.15, 2.65) {};
\node[activepa] (tx2) at ( 1.35, 2.65) {};
\node[activepa] (rx1) at (-0.85, 1.70) {};
\node[activepa] (rx2) at ( 1.65, 1.70) {};

% =====================================================
% One Full-Duplex User Terminal
% =====================================================
\node[userFD] (u) at (0.35, -1.30) {U};
\node[below=2pt, align=center, font=\scriptsize] at (u.south)
{Full-duplex user terminal};

\draw[gray!70, thin] (u.east) -- ++(0.38, 0.12);
\node[anchor=west, align=left, font=\scriptsize, text=gray!70]
at (0.82, -1.12)
{Separate TX/RX chains local SI cancellation};

% =====================================================
% Signal Paths
% =====================================================
% FullPASS-side distributed SI coupling
\draw[->, dashed, thick, red!70]
(tx1.south) -- node[midway, left=-1pt, font=\scriptsize] {SI} (rx1.north);
\draw[->, dashed, thick, red!70] (tx2.south) -- (rx2.north);
\draw[->, dashed, thick, red!70] (tx1.south) -- (rx2.north);
\draw[->, dashed, thick, red!70] (tx2.south) -- (rx1.north);

% Downlink: FullPASS TX waveguide to the same terminal
\draw[->, dashed, thick, blue!70]
(tx1.south) -- (u.north west);
\draw[->, dashed, thick, blue!70]
(tx2.south) -- node[midway, right=1pt, font=\scriptsize] {DL} (u.north east);

% Uplink: the same terminal to the FullPASS RX waveguide
\draw[->, dashed, thick, orange!90!black]
(u.north west) -- (rx1.south);
\draw[->, dashed, thick, orange!90!black]
(u.north east) -- node[midway, right=1pt, font=\scriptsize] {UL} (rx2.south);

% =====================================================
% Dimensions
% =====================================================
\draw[<->, dim] (TXleft |- 0, 2.95) -- (TXright |- 0, 2.95);
\draw[dim, thin] (TXleft) -- ++(0, 0.3);
\draw[dim, thin] (TXright) -- ++(0, 0.3);
\node[above, font=\scriptsize] at (TXleft |- 0, 2.85) {$0$};
\node[above, font=\scriptsize] at (TXright |- 0, 2.85) {$L$};

\draw[<->, dim] (2.65, 1.70) -- (2.85, 2.65);
\node[right, font=\scriptsize] at (2.75, 2.175) {$D$};

\draw[<->, dim] (-3.1, 2.25) -- (-3.1, -1.30);
\node[left, font=\scriptsize] at (-3.1, 0.48) {$H$};

% =====================================================
% Legend
% =====================================================
\node[
draw=gray!40, fill=white, rounded corners=3pt,
thick, inner sep=6pt
] at (0.0, -3.5) {
\begin{tikzpicture}
\node[pa] (p1) at (0,0) {};
\node[right=2pt, font=\scriptsize] at (p1.east) {Potential PA};

```
\node[activepa] (p2) at (2.4,0) {};
\node[right=2pt, font=\scriptsize] at (p2.east) {Activated PA};

\draw[->, dashed, thick, blue!70] (4.8, 0.1) -- (5.4, 0.1);
\draw[->, dashed, thick, orange!90!black] (4.8, -0.1) -- (5.4, -0.1);
\node[right=2pt, font=\scriptsize] at (5.4,0)
  {DL/UL desired links};

\draw[->, dashed, thick, red!70] (7.9,0) -- (8.5,0);
\node[right=2pt, font=\scriptsize] at (8.5,0)
  {FullPASS-side SI};
```

\end{tikzpicture}
};

\end{tikzpicture}}
\caption{Dual-waveguide FullPASS system model with candidate and activated
\acp{PE} on the \ac{TX} and \ac{RX} waveguides and one full-duplex user
terminal that simultaneously receives the downlink signal and transmits the
uplink signal over the same time--frequency resource.}
\label{fig:system_model_pass_selection}
\end{figure}
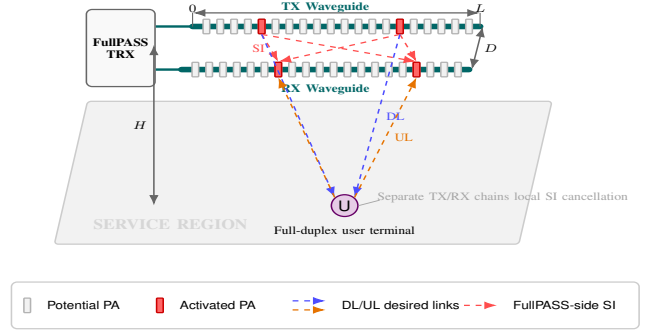

We consider the proposed dual-waveguide FullPASS architecture consisting of
two parallel dielectric waveguides, as shown in
Fig.~\ref{fig:system_model_pass_selection}. The upper waveguide serves as the
\ac{TX} waveguide, while the lower waveguide serves as the \ac{RX} waveguide.
The FullPASS transceiver simultaneously communicates with one full-duplex user
terminal over the same time--frequency resource: the \ac{TX} waveguide
delivers the downlink signal, whereas the \ac{RX} waveguide collects the
uplink signal transmitted by the same terminal. The two waveguides extend
across the served region and are separated by a fixed lateral distance $D$.
The use of dielectric waveguides with localized \acp{PE} follows the
\ac{PASS} operating principle introduced and experimentally demonstrated in prior
work \cite{fukuda2022pinching}.

Activating a \ac{PE} locally perturbs the waveguide and enables coupling
between the guided mode and free space. Thus, the selected \acp{PE} determine
both the desired \ac{DL} and \ac{UL} channels and the cross-waveguide \ac{SI} coupling. The user terminal employs separate transmit and receive \ac{RF} chains
with local passive isolation and conventional analog and digital
\ac{SI} cancellation. This work focuses on propagation-domain
\ac{SI} suppression at the FullPASS transceiver. Therefore, any
terminal-side \ac{SI} leakage is assumed to be independently
controlled, unaffected by the selected FullPASS \acp{PE}, and
absorbed into a fixed effective \ac{DL} noise-plus-interference power.

The \ac{TX} waveguide contains $N_{\mathrm{TX}}$ candidate \acp{PE}
$d_i^{\mathrm{TX}}$, $i=1,\ldots,N_{\mathrm{TX}}$, and the \ac{RX} waveguide
contains $N_{\mathrm{RX}}$ candidate \acp{PE}
$d_j^{\mathrm{RX}}$, $j=1,\ldots,N_{\mathrm{RX}}$. We adopt a Cartesian coordinate system where both parallel waveguides reside in the horizontal plane $z=0$. The \ac{RX} waveguide is aligned with the $x$-axis at $y=0$, while the \ac{TX} waveguide is shifted parallel to it at a fixed lateral distance $y=D$. Thus, the discrete position vectors of the $i$th \ac{TX} and $j$th \ac{RX} candidate locations are given by
\begin{equation}
\mathbf{r}_i^{(\mathrm{TX})}
=
[d_i^{\mathrm{TX}},D,0]^T,
\qquad
\mathbf{r}_j^{(\mathrm{RX})}
=
[d_j^{\mathrm{RX}},0,0]^T.
\end{equation}
Both waveguide ports are referenced at $x=0$, so that
$d_i^{\rm TX}$ and $d_j^{\rm RX}$ also represent the corresponding guided
propagation distances from the \ac{TX} feed and to the \ac{RX} chain,
respectively.

The practical FullPASS model developed below builds on two established
\ac{PASS} modeling components. First, we adopt the proportional-power
coupling model in \cite{wang2025modeling}, in which identical activated
\acp{PE} couple a fixed fraction of the guided power remaining in the
waveguide. Second, we follow the attenuation-aware channel decomposition in
\cite{xu2026inwaveguide}, which accounts for free-space propagation together
with in-waveguide propagation phase and attenuation. We apply these modeling
components to both the transmit and receive waveguides and extend them to
describe the pairwise transmit-to-receive \ac{SI} coupling in the proposed
dual-waveguide FullPASS architecture.

\subsection{Binary Activation and Cumulative Pass-Through Loss}

We define binary activation variables
\begin{equation}
a_i^{\mathrm{TX}}\in\{0,1\},
\qquad
i=1,\ldots,N_{\mathrm{TX}},
\end{equation}
and
\begin{equation}
a_j^{\mathrm{RX}}\in\{0,1\},
\qquad
j=1,\ldots,N_{\mathrm{RX}}.
\end{equation}
Here, $a_i^{\mathrm{TX}}=1$ means that the $i$th \ac{TX} candidate \ac{PE} is
activated, while $a_i^{\mathrm{TX}}=0$ means that it is inactive. The \ac{RX}
activation variables are defined similarly.

We assume that all activated \ac{TX} \acp{PE} have the same amplitude
coupling coefficient $\delta_{\mathrm{TX}}$ and pass-through coefficient
$\tau_{\mathrm{TX}}$, and that all activated \ac{RX} \acp{PE} have the same amplitude coupling coefficient $\delta_{\mathrm{RX}}$ and pass-through
coefficient $\tau_{\mathrm{RX}}$. The coefficient $\delta_s$ scales the signal amplitude coupled between the
waveguide and free space on side $s\in\{\mathrm{TX},\mathrm{RX}\}$, while
$\tau_s$ scales the signal amplitude that remains guided and continues along
the waveguide after an activated \ac{PE}. We assume that the coupling coefficient $\delta_s$ and the pass-through
coefficient $\tau_s$ are real and nonnegative, with
\begin{equation}
\delta_s^2+\tau_s^2=1,
\qquad s\in\{\mathrm{TX},\mathrm{RX}\},
\label{eq:power_conservation}
\end{equation}
where $\delta_s^2$ denotes the fraction of power coupled between the waveguide
and free space, while $\tau_s^2$ denotes the fraction of power that remains in
the waveguide after an activated \ac{PE}. This relation represents lossless
local power conservation at each activated \ac{PE}. Any phase shift introduced
by an activated \ac{PE} is assumed to be identical across all candidate
locations and is absorbed into the propagation coefficients defined below.

\begin{figure}[t!]
\centering
\resizebox{0.95\columnwidth}{!}{\begin{tikzpicture}[>=Latex, font=\small\sffamily]
\draw[line width=4pt, teal!80!black] (0,0) -- (7,0);
\node[
draw=gray!80!black,
fill=gray!5,
thick,
rounded corners=2pt,
minimum width=0.8cm,
minimum height=1cm,
align=center,
font=\bfseries\scriptsize
] (trx) at (-0.6,0) {TX};
\draw[line width=1.5pt, teal!80!black] (trx.east) -- (0,0);

% uniformly spaced candidate locations (all inactive first)
\foreach \x in {0.5,1.0,1.5,2.0,2.5,3.0,3.5,4.0,4.5,5.0,5.5,6.0,6.5} {
    \node[
        draw=gray!60,
        fill=gray!10,
        thick,
        rounded corners=0.5pt,
        minimum width=0.12cm,
        minimum height=0.32cm,
        inner sep=0pt
    ] at (\x,0) {};
}

% active pinching points drawn on top
\foreach \x/\lbl in {1.5/1, 3.5/2, 5.5/3} {
    \node[
        draw=red!80!black,
        fill=red!60,
        thick,
        rounded corners=0.5pt,
        minimum width=0.12cm,
        minimum height=0.32cm,
        inner sep=0pt
    ] (pa\lbl) at (\x,0) {};
    \draw[->, dashed, thick, red!70]
        (pa\lbl.south) -- ++(0,-0.6)
        node[below, font=\scriptsize] {Coupling};
}

\draw[->, line width=2pt, blue!80!black]
    (0.2,0.4) -- (1.3,0.4)
    node[midway, above, font=\scriptsize] {$p_1^{\rm TX}$};

\draw[->, line width=1.5pt, blue!80!black, opacity=0.8]
    (1.7,0.4) -- (3.3,0.4)
    node[midway, above, font=\scriptsize] {$p_2^{\rm TX}=p_1^{\rm TX}\tau_{\rm TX}$};

\draw[->, line width=1pt, blue!80!black, opacity=0.6]
    (3.7,0.4) -- (5.3,0.4)
    node[midway, above, font=\scriptsize] {$p_3^{\rm TX}=p_2^{\rm TX}\tau_{\rm TX}$};

\draw[->, line width=0.5pt, blue!80!black, opacity=0.4]
    (5.7,0.4) -- (7.0,0.4)
    node[midway, above, font=\scriptsize] {$p_4^{\rm TX}=p_3^{\rm TX}\tau_{\rm TX}$};

% legend
\node[
    draw=red!80!black,
    fill=red!60,
    thick,
    rounded corners=0.5pt,
    minimum width=0.12cm,
    minimum height=0.32cm,
    inner sep=0pt
] at (5.8,-1.4) {};
\node[anchor=west, font=\scriptsize] at (6.0,-1.4) {active pinching point};

\node[
    draw=gray!60,
    fill=gray!10,
    thick,
    rounded corners=0.5pt,
    minimum width=0.12cm,
    minimum height=0.32cm,
    inner sep=0pt
] at (5.8,-1.85) {};
\node[anchor=west, font=\scriptsize] at (6.0,-1.85) {inactive candidate location};

\end{tikzpicture}}
\caption{Illustration of cascaded power flow and cumulative pass-through loss along a series-fed waveguide, shown for the \ac{TX} side. The same
cumulative-loss weighting is used for the \ac{RX} waveguide.}
\label{fig:cascaded_power}
\end{figure}
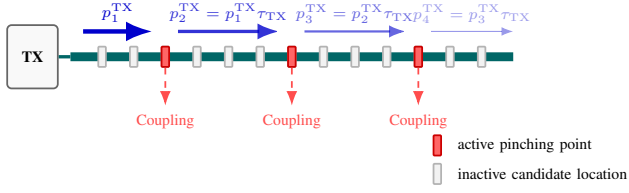

An active \ac{PE} reduces the guided-signal amplitude available at downstream
candidate locations. We capture this activation-dependent cumulative pass-through loss through
the effective weights
\begin{equation}
y_i^{\mathrm{TX}}
=
a_i^{\mathrm{TX}}
\tau_{\mathrm{TX}}^{\sum_{\ell<i}a_\ell^{\mathrm{TX}}},
\qquad
y_j^{\mathrm{RX}}
=
a_j^{\mathrm{RX}}
\tau_{\mathrm{RX}}^{\sum_{\ell<j}a_\ell^{\mathrm{RX}}}.
\label{eq:prefix_weights}
\end{equation}
Thus, an inactive candidate has zero weight, whereas an active candidate is
attenuated by the pass-through factors of upstream active \acp{PE}. This
activation-dependent attenuation is distinct from the continuous guided-wave
loss included in the channel coefficients below. The corresponding
cascaded power-flow interpretation is illustrated in
Fig.~\ref{fig:cascaded_power}.

\subsection{Complex Channel Coefficients}
\label{subsec:propagation_coefficients}

We use a geometry-driven line-of-sight model in which each complex channel
coefficient captures free-space spherical spreading and phase, guided-wave
phase accumulation, and continuous guided-wave attenuation. Together with the local coupling coefficients and the activation-dependent
pass-through factors, these propagation coefficients form a system-level
channel model for discrete FullPASS geometry selection. Full-wave
electromagnetic modeling would be required to characterize mutual coupling,
local loading, radiation patterns, and hardware-specific scattering effects;
such validation is left for future work.

Let $\lambda_0$ denote the carrier wavelength and
$k_0=2\pi/\lambda_0$ the free-space wavenumber. The guided-wave phase
accumulated over distance $d$ is $k_0n_e d$, where $n_e$ is the effective
refractive index. The continuous guided-wave attenuation coefficients
$\alpha_{\rm TX}$ and $\alpha_{\rm RX}$ are expressed in $\mathrm{dB/m}$.

The full-duplex user terminal is located in the service region at height $H$
below the waveguide plane. Its transmit and receive ports are assumed
co-located relative to the considered FullPASS geometry. We write its position
as
\begin{equation}
\mathbf r_u
=
[x_u,y_u,-H]^T,
\label{eq:fd_user_coordinate}
\end{equation}
where $(x_u,y_u)$ denotes the horizontal terminal location and
$H$ is the perpendicular distance from the waveguide plane.

For the downlink, the distance to the $i$th \ac{TX} candidate is
\begin{equation}
\begin{aligned}
R_{{\rm DL},i}
&=
\left\|\mathbf r_u-\mathbf r_i^{(\mathrm{TX})}\right\|
\\
&=
\sqrt{
\left(x_u-d_i^{\mathrm{TX}}\right)^2
+
\left(y_u-D\right)^2
+
H^2
},
\end{aligned}
\label{eq:R_dl_i}
\end{equation}
and the corresponding complex \ac{DL} channel coefficient is
\begin{equation}
g_{{\rm DL},i}
=
10^{-\alpha_{\rm TX}d_i^{\rm TX}/20}
\frac{\lambda_0}{4\pi R_{{\rm DL},i}}
e^{-jk_0\left(R_{{\rm DL},i}+n_e d_i^{\rm TX}\right)}.
\label{eq:g_dl}
\end{equation}

For the uplink, the distance to the $j$th \ac{RX} candidate is
\begin{equation}
R_{{\rm UL},j}
=
\left\|\mathbf r_u-\mathbf r_j^{(\mathrm{RX})}\right\|
=
\sqrt{
\left(x_u-d_j^{\mathrm{RX}}\right)^2
+
y_u^2
+
H^2
},
\label{eq:R_ul_j}
\end{equation}
and the corresponding complex \ac{UL} channel coefficient is
\begin{equation}
g_{{\rm UL},j}
=
10^{-\alpha_{\rm RX}d_j^{\rm RX}/20}
\frac{\lambda_0}{4\pi R_{{\rm UL},j}}
e^{-jk_0\left(R_{{\rm UL},j}+n_e d_j^{\rm RX}\right)}.
\label{eq:g_ul}
\end{equation}

The distance between \ac{TX} candidate $i$ and \ac{RX} candidate $j$ is
\begin{equation}
r_{j,i}
=
\left\|\mathbf r_j^{(\mathrm{RX})}-\mathbf r_i^{(\mathrm{TX})}\right\|
=
\sqrt{\left(d_j^{\mathrm{RX}}-d_i^{\mathrm{TX}}\right)^2+D^2}.
\label{eq:r_ji}
\end{equation}
This \ac{SI} distance is determined by the two waveguide coordinates and does
not depend on the user coordinates. The corresponding complex distributed \ac{SI} channel coefficient is
\begin{equation}
\begin{aligned}
G_{j,i}^{\rm SI}
={}&
10^{-\left(\alpha_{\rm RX}d_j^{\rm RX}
+\alpha_{\rm TX}d_i^{\rm TX}\right)/20}
\frac{\lambda_0}{4\pi r_{j,i}}
\\[-1mm]
&\times
e^{-jk_0\left[r_{j,i}
+n_e\left(d_i^{\rm TX}+d_j^{\rm RX}\right)\right]}.
\end{aligned}
\label{eq:g_si}
\end{equation}

In the following sections, we assume perfect \ac{CSI} at the FullPASS
transceiver to isolate the discrete \ac{PE}-selection problem. This assumption
is more reasonable for the FullPASS-side \ac{SI} channel than for the desired
user links: for a fixed activation pattern and hardware configuration, its
dominant local coupling paths vary slowly and can be re-estimated after the
activation is selected. The model nevertheless neglects multipath-induced and
time-varying \ac{SI} components. Robust selection under imperfect desired-link
\ac{CSI}, \ac{SI}-model mismatch, and position uncertainty is left for future
work.

\subsection{Effective Channels and Received Powers}

Using the effective activation weights in
\eqref{eq:prefix_weights}, the effective \ac{DL} and \ac{UL} channels are
\begin{equation}
H_{\mathrm{DL}}
=
\delta_{\mathrm{TX}}
\sum_{i=1}^{N_{\mathrm{TX}}}
g_{{\rm DL},i}
y_i^{\mathrm{TX}},
\label{eq:H_DL_selection}
\end{equation}
and
\begin{equation}
H_{\mathrm{UL}}
=
\delta_{\mathrm{RX}}
\sum_{j=1}^{N_{\mathrm{RX}}}
g_{{\rm UL},j}
y_j^{\mathrm{RX}}.
\label{eq:H_UL_selection}
\end{equation}
The aggregate cross-waveguide \ac{SI} channel is the coherent double sum
\begin{equation}
H_{\mathrm{SI}}
=
\delta_{\mathrm{RX}}\delta_{\mathrm{TX}}
\sum_{j=1}^{N_{\mathrm{RX}}}
\sum_{i=1}^{N_{\mathrm{TX}}}
y_j^{\mathrm{RX}}
G_{j,i}^{\rm SI}
y_i^{\mathrm{TX}}.
\label{eq:H_SI_selection}
\end{equation}

For compactness, define
\begin{equation}
\begin{aligned}
\mathbf g_{\rm DL}
&=
[g_{{\rm DL},1},\ldots,g_{{\rm DL},N_{\rm TX}}]^T,
\\
\mathbf g_{\rm UL}
&=
[g_{{\rm UL},1},\ldots,g_{{\rm UL},N_{\rm RX}}]^T,
\end{aligned}
\end{equation}
\begin{equation}
\begin{aligned}
\mathbf y_{\rm TX}
&=
[y_1^{\rm TX},\ldots,y_{N_{\rm TX}}^{\rm TX}]^T,
\\
\mathbf y_{\rm RX}
&=
[y_1^{\rm RX},\ldots,y_{N_{\rm RX}}^{\rm RX}]^T,
\end{aligned}
\end{equation}
and
\begin{equation}
[\mathbf G_{\rm SI}]_{j,i}=G_{j,i}^{\rm SI}.
\end{equation}
Because the activation vectors are real-valued and the complex propagation
phases are already included in $\mathbf g_{\rm DL}$, $\mathbf g_{\rm UL}$,
and $\mathbf G_{\rm SI}$, the ordinary transpose $(\cdot)^T$ is used. The scalar effective channel links are therefore modeled as linear directional combinations rather than complex inner products. With these definitions,
\begin{equation}
H_{\rm DL}
=
\delta_{\rm TX}
\mathbf g_{\rm DL}^{T}\mathbf y_{\rm TX},
\label{eq:H_DL_vector}
\end{equation}
\begin{equation}
H_{\rm UL}
=
\delta_{\rm RX}
\mathbf g_{\rm UL}^{T}\mathbf y_{\rm RX},
\label{eq:H_UL_vector}
\end{equation}
and
\begin{equation}
H_{\rm SI}
=
\delta_{\rm RX}\delta_{\rm TX}
\mathbf y_{\rm RX}^{T}
\mathbf G_{\rm SI}
\mathbf y_{\rm TX}.
\label{eq:H_SI_vector}
\end{equation}
Equivalently, \eqref{eq:H_SI_vector} is the matrix form of the double sum in
\eqref{eq:H_SI_selection}. Let $P_{\mathrm{TX}}$ denote the transmit power of
the FullPASS \ac{TX} chain and let $P_{\mathrm{U}}$ denote the fixed transmit
power of the full-duplex user terminal. The received desired-link powers are
\begin{equation}
P_{\mathrm{DL}}
=
P_{\mathrm{TX}}|H_{\mathrm{DL}}|^2,
\label{eq:P_DL}
\end{equation}
and
\begin{equation}
P_{\mathrm{UL}}
=
P_{\mathrm{U}}|H_{\mathrm{UL}}|^2.
\label{eq:P_UL}
\end{equation}
The coupled \ac{SI} power at the FullPASS \ac{RX} chain input is
\begin{equation}
P_{\mathrm{SI}}
=
P_{\mathrm{TX}}|H_{\mathrm{SI}}|^2.
\label{eq:P_SI}
\end{equation}

% ======================================================
\section{Problem Formulation}
\label{sec:problem}
% ======================================================

This section formulates the discrete \ac{PE}--selection problem introduced by
the channel model of Section~\ref{sec:system_model}, under a prescribed
\ac{SI}-leakage requirement.

Building on the effective channels and received powers derived in
Section~\ref{sec:system_model}, we now cast joint transmit--receive
pinching-element selection as a discrete optimization problem. We define this
problem in three steps: we first specify the binary design variables and their
admissible set, then formulate the desired-link objective and identify the
structure we exploit for selection, and finally impose the \ac{SI}-leakage
constraint that couples the two waveguides. Assembling these three components
yields the selection problem solved in the remainder of the paper.

The design variables are the binary activation vectors $\mathbf a_{\rm TX}$ and
$\mathbf a_{\rm RX}$ defined in Section~\ref{sec:system_model}, whose entries
select the active transmit and receive \acp{PE}. We restrict these vectors to
the admissible set
\begin{align}
\mathcal A
=
\Big\{
(\mathbf a_{\rm TX},\mathbf a_{\rm RX})&:\,
a_i^{\rm TX}\in\{0,1\},\quad
a_j^{\rm RX}\in\{0,1\},
\nonumber\\[-0.2em]
&
\sum_i a_i^{\rm TX}\ge 1,\quad
\sum_j a_j^{\rm RX}\ge 1
\Big\}.
\label{eq:activation_set}
\end{align}
The nonempty-side conditions in $\mathcal A$ exclude trivial activations with no
desired transmit or receive link, ensuring that every candidate solution
supports at least one downlink and one uplink path.

As the objective, we adopt the bidirectional desired-link sum spectral
efficiency. Let $\sigma_{\rm DL}^2$ and $\sigma_{\rm UL}^2$ denote the fixed
effective noise-plus-interference powers at the full-duplex user terminal and
the FullPASS receiver, respectively. With the received desired-link powers
$P_{\rm DL}$ and $P_{\rm UL}$ in \eqref{eq:P_DL} and \eqref{eq:P_UL}, this sum
spectral efficiency is
\begin{align}
R_{\rm sum}
=
&
\log_2\left(1+\frac{P_{\rm DL}}{\sigma_{\rm DL}^2}\right)
\nonumber\\[-0.15em]
&+
\log_2\left(1+\frac{P_{\rm UL}}{\sigma_{\rm UL}^2}\right).
\label{eq:sum_rate_original}
\end{align}
Two structural properties of \eqref{eq:sum_rate_original} shape the selection
problem. First, the objective is separable across the two waveguides: through
\eqref{eq:H_DL_vector} and \eqref{eq:H_UL_vector}, the downlink term depends
only on $\mathbf a_{\rm TX}$ and the uplink term only on $\mathbf a_{\rm RX}$,
so the transmit and receive selections are coupled solely through the
\ac{SI}-leakage constraint introduced next. Second, each term is a strictly
increasing function of the corresponding received power, and hence of the
desired channel magnitude $|H_{\rm DL}|$ or $|H_{\rm UL}|$; the additive noise
powers act only as fixed scalings and do not affect which activation maximizes
a given term. Consequently, maximizing \eqref{eq:sum_rate_original} amounts to
jointly strengthening the two desired channel magnitudes subject to the
\ac{SI}-leakage limit, which is the view exploited in the selection framework
of Section~\ref{sec:proposed_framework}.

As the constraint, we treat the cross-waveguide \ac{SI} as a leakage budget
rather than as an additive noise term. Accordingly, the coupled \ac{SI} power
$P_{\rm SI}$ in \eqref{eq:P_SI} does not enter the uplink-rate denominator of
\eqref{eq:sum_rate_original}; instead, the transmit and receive pinching
elements are selected to keep $P_{\rm SI}$ below a prescribed limit that
subsequent analog and digital cancellation stages can handle. Using superscript
`pr' for quantities evaluated under the practical FullPASS model of
Section~\ref{sec:system_model}, and writing $P_{\rm SI,max}^{\rm pr}$ for this
limit, we impose the hard \ac{SI}-leakage constraint
$P_{\rm TX}|H_{\rm SI}|^2\le P_{\rm SI,max}^{\rm pr}$.

Combining the objective, the admissible set $\mathcal A$, and the
\ac{SI}-leakage constraint yields the discrete FullPASS selection problem
\begin{equation}
\begin{aligned}
\max_{(\mathbf a_{\rm TX},\mathbf a_{\rm RX})\in\mathcal A}
\quad &
\log_2\left(1+\frac{P_{\rm DL}}{\sigma_{\rm DL}^2}\right)
+
\log_2\left(1+\frac{P_{\rm UL}}{\sigma_{\rm UL}^2}\right)
\\
\text{s.t.}\quad &
P_{\rm TX}|H_{\rm SI}|^2
\le
P_{\rm SI,max}^{\rm pr}.
\end{aligned}
\label{eq:practical_hard_problem}
\end{equation}

Problem \eqref{eq:practical_hard_problem} is nonconvex and combinatorial. The
activation variables are binary; the practical weights in
\eqref{eq:prefix_weights} depend on the number and order of upstream activated
\acp{PE}; the desired-link powers are squared magnitudes of coherent
complex-valued sums; and the \ac{SI} channel in \eqref{eq:H_SI_vector} couples
the \ac{TX} and \ac{RX} activation vectors bilinearly. Exhaustive search is
therefore limited to moderate candidate grids, motivating the scalable
candidate-generation and refinement methods developed next.

% ======================================================
\section{Proposed Two-Stage Selection Framework}
\label{sec:proposed_framework}
% ======================================================

We use a two-stage framework to address the coupled binary selection problem in
\eqref{eq:practical_hard_problem}. Stage~1 searches for activation patterns
$(\mathbf a_{\rm TX},\mathbf a_{\rm RX})\in\mathcal A$ that increase the
objective \eqref{eq:sum_rate_original} while satisfying the practical-model
leakage constraint $P_{\rm TX}|H_{\rm SI}|^2\le P_{\rm SI,max}^{\rm pr}$, using a
simplified propagation model and a calibrated guard margin to generate binary
trial patterns. Stage~2 further refines the resulting pattern by local search
over the same feasible set, again maximizing \eqref{eq:sum_rate_original} under
the practical model.

% ======================================================
\subsection{Simplified In-Waveguide Propagation Model}
\label{subsec:equal_feed_surrogate_model}
% ======================================================

For trial-pattern generation, we omit only the activation-dependent
cumulative pass-through loss. Thus, each active candidate has unit selection weight,
whereas continuous in-waveguide attenuation, geometry-dependent phases, and distributed
\ac{SI} coupling remain in the model. The resulting patterns are subsequently
screened and refined under the practical model.
\begin{equation}
y_i^{\rm TX}=a_i^{\rm TX},
\qquad
y_j^{\rm RX}=a_j^{\rm RX}.
\label{eq:ideal_y_equals_a}
\end{equation}
This idealized unit-weight approximation, which temporarily ignores cumulative pass-through loss, is used only for trial pattern generation. 

Substituting \eqref{eq:ideal_y_equals_a} into
\eqref{eq:H_DL_vector}--\eqref{eq:H_SI_vector}, and using the superscript
`sm' to denote quantities evaluated under the simplified model, the
corresponding downlink and uplink channels are
\begin{equation}
H_{\rm DL}^{\rm sm}
=
\delta_{\rm TX}\mathbf g_{\rm DL}^{T}\mathbf a_{\rm TX},
\qquad
H_{\rm UL}^{\rm sm}
=
\delta_{\rm RX}\mathbf g_{\rm UL}^{T}\mathbf a_{\rm RX},
\label{eq:H_desired_equal_feed}
\end{equation}
and the corresponding cross-waveguide \ac{SI} channel is
\begin{equation}
H_{\rm SI}^{\rm sm}
=
\delta_{\rm RX}\delta_{\rm TX}
\mathbf a_{\rm RX}^{T}\mathbf G_{\rm SI}\mathbf a_{\rm TX}.
\label{eq:H_SI_equal_feed}
\end{equation}
The corresponding downlink and uplink received powers are
\begin{equation}
P_{\rm DL}^{\rm sm}
=
P_{\rm TX}|H_{\rm DL}^{\rm sm}|^2,
\qquad
P_{\rm UL}^{\rm sm}
=
P_{\rm U}|H_{\rm UL}^{\rm sm}|^2.
\label{eq:P_desired_equal_feed}
\end{equation}

The simplified model uses the same desired-link objective \eqref{eq:sum_rate_original}, with $P_{\rm DL}$ and $P_{\rm UL}$ replaced by $P_{\rm DL}^{\rm sm}$ and $P_{\rm UL}^{\rm sm}$, respectively.

\subsection{Guard-Margin Calibration for the Simplified \acs{SI} Constraint}
\label{subsec:guard_margin}

Because replacing the activation weights of practical model with the unit weights of
the simplified model alters the coherent addition of the individual \ac{SI}
coupling paths, the practical-model \ac{SI}-leakage power can occasionally
exceed the simplified prediction. We therefore introduce a calibrated guard margin. Specifically, we define the model mismatch as
\begin{equation}
\Delta_{\rm SI} = P_{\rm SI}^{\rm pr,dBm} - P_{\rm SI}^{\rm sm,dBm}.
\label{eq:delta_si_def}
\end{equation}

Here, $P_{\rm SI}^{\rm pr,dBm}$ and $P_{\rm SI}^{\rm sm,dBm}$ denote the
practical-model and simplified-model \ac{SI}-leakage powers, respectively,
expressed in dBm.

We then apply a non-negative guard margin, $\Delta_{\rm guard}\ge0$, to establish a tightened threshold for the simplified model:
\begin{equation}
P_{\rm SI,max}^{\rm sm,dBm} = P_{\rm SI,max}^{\rm pr,dBm} - \Delta_{\rm guard}.
\label{eq:guarded_threshold}
\end{equation}
The trial pattern generation constraint evaluated under the simplified model is therefore
\begin{equation}
P_{\rm TX}|H_{\rm SI}^{\rm sm}|^2 \le P_{\rm SI,max}^{\rm sm}.
\label{eq:equal_feed_hard_si_constraint}
\end{equation}
Consequently, any trial pattern satisfying \eqref{eq:equal_feed_hard_si_constraint} inherently satisfies the practical-model \ac{SI}-leakage constraint in \eqref{eq:practical_hard_problem} provided $\Delta_{\rm SI}\le\Delta_{\rm guard}$.

We calibrate $\Delta_{\rm guard}$ offline using randomly generated nonempty
activation patterns. The \ac{TX} and \ac{RX} activation cardinalities are
sampled uniformly over their admissible ranges, and the active locations are
sampled uniformly for each selected cardinality. Letting $F_{\Delta}$ denote the empirical cumulative distribution function of $\Delta_{\rm SI}$, we set the margin for a target mismatch probability $\epsilon$ as
\begin{equation}
\Delta_{\rm guard}
=
\max\!\left\{0,F_{\Delta}^{-1}(1-\epsilon)\right\}.
\label{eq:guard_quantile}
\end{equation}

This guard is used strictly during trial pattern generation; every final activation is explicitly checked against the practical limit in the constraint of \eqref{eq:practical_hard_problem}.

% ======================================================
\subsection{Stage 1: Phase-Anchored \acs{SOCP} Initialization}
\label{subsec:phase_socp_surrogate}
% ======================================================

Stage~1 searches for initial activation patterns
$(\mathbf a_{\rm TX},\mathbf a_{\rm RX})\in\mathcal A$ that increase the
objective \eqref{eq:sum_rate_original} subject to the practical-model leakage
constraint in \eqref{eq:practical_hard_problem}, using the simplified model in
Section~\ref{subsec:equal_feed_surrogate_model} for candidate generation. The binary activation
variables are first relaxed to continuous values between zero and one. The
\ac{TX} and \ac{RX} variables are optimized alternately: when optimizing one
side, the variables on the other side are kept fixed. We refer to each such
single-waveguide optimization step -- \ac{TX} or \ac{RX} -- as a block. For each phase
anchor and block update, the relaxed solution is deterministically rounded
into binary trial patterns. These trial patterns are evaluated under the practical model, and the best pattern satisfying the \ac{SI}-leakage constraint in
\eqref{eq:practical_hard_problem} is accepted only if it improves the incumbent sum spectral efficiency \eqref{eq:sum_rate_original}. Stage~1 terminates when no
accepted block update improves the incumbent. Its resulting binary activation
is then passed to Stage~2 for practical-model \ac{ABLS} refinement, as
summarized in Algorithm~\ref{alg:phase_quant_socp}.

% ======================================================
\subsubsection{Alternating TX and RX SOCP Subproblems}
\label{subsec:phase_quantized_socp}
% ======================================================

Consider one of the alternating subproblems, in which the activation variables
on one waveguide are optimized while those on the other waveguide are fixed.
The uplink or downlink channel is then a complex-valued linear function of the relaxed
activation variables. Directly maximizing the magnitude of this channel is
nonconvex. The phase of the uplink or downlink complex channel is generally
unknown before solving the subproblem. To obtain convex subproblems, we
therefore test a finite set of $Q_\phi$ evenly spaced reference phases, which
we call phase anchors,
\begin{equation}
\phi_q=\frac{2\pi q}{Q_\phi},
\qquad q=0,\ldots,Q_\phi-1.
\label{eq:phase_grid}
\end{equation}
For each phase anchor, the channel is rotated so that this phase lies on the
positive real axis, and its real part is maximized.

For any complex channel value $h$, one of these phase anchors is close to
$\angle h$. Rotating $h$ by that phase makes its real part close to its
magnitude. More precisely,
\begin{equation}
\max_q \Re\left\{e^{-j\phi_q}h\right\}
\ge
|h|\cos\left(\frac{\pi}{Q_\phi}\right).
\label{eq:phase_bound}
\end{equation}
Thus, maximizing the best phase-rotated real part provides a controlled
approximation of maximizing $|h|$; increasing $Q_\phi$ improves this
approximation.

We apply this phase-based approximation separately to the \ac{TX} and \ac{RX}
updates. When the \ac{RX} activation is fixed, the uplink received power is
constant, so the \ac{TX} update only needs to improve the downlink channel
while satisfying the simplified-model \ac{SI}-leakage constraint. Similarly, when
the \ac{TX} activation is fixed, the \ac{RX} update improves the uplink
channel. The following \ac{SOCP} formulations implement these two alternating
updates.

We first describe the \ac{TX} update. The current binary \ac{RX} activation
pattern $\mathbf a_{\rm RX}$ is kept fixed, whereas the \ac{TX} activation
vector is temporarily relaxed from binary values to
$\mathbf u_{\rm TX}\in[0,1]^{N_{\rm TX}}$. Under the simplified model, the aggregate \ac{SI} channel in
\eqref{eq:H_SI_equal_feed} is linear in $\mathbf u_{\rm TX}$ when
$\mathbf a_{\rm RX}$ is fixed. In
particular, for each \ac{TX} candidate $i$, the quantity
$\bar c_i^{\rm TX}$ collects the total \ac{SI} coupling from that candidate to
all currently active \ac{RX} candidates. It includes the corresponding
\ac{TX}--\ac{RX} coupling coefficients and the fixed \ac{RX}-side coupling
factor:
\begin{equation}
\bar c_i^{\rm TX}
=
\delta_{\rm RX}
\sum_{j=1}^{N_{\rm RX}}
a_j^{\rm RX}G_{j,i}^{\rm SI},
\qquad i=1,\ldots,N_{\rm TX}.
\label{eq:ctx_bar_def}
\end{equation}
Thus, after fixing $\mathbf a_{\rm RX}$, the aggregate \ac{SI} amplitude is
determined by the complex scalar
$\bar{\mathbf c}_{\rm TX}^{T}\mathbf u_{\rm TX}$. The coefficient
$\eta_{\rm TX}$ below combines the \ac{TX} power, the local \ac{TX} coupling
factor, and the guarded simplified-model \ac{SI} threshold:
\begin{equation}
\eta_{\rm TX}
\triangleq
\frac{P_{\rm TX}|\delta_{\rm TX}|^2}
{P_{\rm SI,max}^{\rm sm}}.
\label{eq:tx_eta_def}
\end{equation}
To express the magnitude of this complex \ac{SI} amplitude using a
second-order-cone constraint, we stack its real and imaginary parts as
\begin{equation}
\mathbf z_{\rm TX}(\mathbf u_{\rm TX})
\triangleq
\begin{bmatrix}
\Re\{\bar{\mathbf c}_{\rm TX}^{T}\mathbf u_{\rm TX}\}\\
\Im\{\bar{\mathbf c}_{\rm TX}^{T}\mathbf u_{\rm TX}\}
\end{bmatrix},
\label{eq:tx_z_def}
\end{equation}
so that $\|\mathbf z_{\rm TX}(\mathbf u_{\rm TX})\|_2$ equals the magnitude of
$\bar{\mathbf c}_{\rm TX}^{T}\mathbf u_{\rm TX}$.

The desired downlink channel is also linear in $\mathbf u_{\rm TX}$. For a
given phase-grid point $\phi_q$, we maximize the real part of the downlink
channel after rotation by $e^{-j\phi_q}$:
\begin{equation}
\psi_{\rm TX}^{(q)}(\mathbf u_{\rm TX})
\triangleq
\Re\left\{
e^{-j\phi_q}
\delta_{\rm TX}\mathbf g_{\rm DL}^{T}\mathbf u_{\rm TX}
\right\}.
\label{eq:tx_phase_link_def}
\end{equation}
For each phase-grid point, the resulting problem maximizes this
phase-anchored downlink projection. Its first constraint imposes the guarded
simplified-model \ac{SI} limit, its second constraint is the continuous
relaxation of the binary \ac{TX} activations, and its final constraint excludes
the all-zero activation:
\begin{subequations}
\label{eq:tx_phase_socp}
\begin{align}
\max_{\mathbf u_{\rm TX}}
\quad &
\psi_{\rm TX}^{(q)}(\mathbf u_{\rm TX})
\label{eq:tx_phase_socp_obj}
\\
\text{s.t.}\quad
&
\sqrt{\eta_{\rm TX}}\,
\|\mathbf z_{\rm TX}(\mathbf u_{\rm TX})\|_2
\le 1,
\label{eq:tx_phase_socp_si}
\\
&
0\le u_i^{\rm TX}\le 1,
\quad i=1,\ldots,N_{\rm TX},
\label{eq:tx_phase_socp_box}
\\
&
\sum_{i=1}^{N_{\rm TX}}u_i^{\rm TX}
\ge 1.
\label{eq:tx_phase_socp_min}
\end{align}
\end{subequations}
For fixed $\phi_q$ and fixed $\mathbf a_{\rm RX}$, the objective is linear,
the \ac{SI} constraint is second-order conic, and the remaining constraints are
linear. Hence, \eqref{eq:tx_phase_socp} is a convex \ac{SOCP}.

The \ac{RX} update is obtained by symmetry from the preceding \ac{TX} update.
Here, the \ac{TX} activation pattern $\mathbf a_{\rm TX}$ is fixed and the
\ac{RX} activation vector is relaxed to
$\mathbf u_{\rm RX}\in[0,1]^{N_{\rm RX}}$. For each \ac{RX} candidate $j$,
the coefficient below aggregates its \ac{SI} coupling from all
currently active \ac{TX} candidates:
\begin{equation}
\bar c_j^{\rm RX}
=
\delta_{\rm TX}
\sum_{i=1}^{N_{\rm TX}}
G_{j,i}^{\rm SI}a_i^{\rm TX},
\qquad j=1,\ldots,N_{\rm RX},
\label{eq:crx_bar_def}
\end{equation}
and the corresponding normalized \ac{SI} coefficient is
\begin{equation}
\eta_{\rm RX}
\triangleq
\frac{P_{\rm TX}|\delta_{\rm RX}|^2}
{P_{\rm SI,max}^{\rm sm}}.
\label{eq:rx_eta_def}
\end{equation}
The real-valued \ac{SI} representation and the phase-anchored uplink
projection are
\begin{equation}
\mathbf z_{\rm RX}(\mathbf u_{\rm RX})
\triangleq
\begin{bmatrix}
\Re\{\bar{\mathbf c}_{\rm RX}^{T}\mathbf u_{\rm RX}\}\\
\Im\{\bar{\mathbf c}_{\rm RX}^{T}\mathbf u_{\rm RX}\}
\end{bmatrix},
\label{eq:rx_z_def}
\end{equation}
and
\begin{equation}
\psi_{\rm RX}^{(q)}(\mathbf u_{\rm RX})
\triangleq
\Re\left\{
e^{-j\phi_q}
\delta_{\rm RX}\mathbf g_{\rm UL}^{T}\mathbf u_{\rm RX}
\right\}.
\label{eq:rx_phase_link_def}
\end{equation}
Accordingly, for each phase-grid point, the relaxed \ac{RX} subproblem is
\begin{subequations}
\label{eq:rx_phase_socp}
\begin{align}
\max_{\mathbf u_{\rm RX}}
\quad &
\psi_{\rm RX}^{(q)}(\mathbf u_{\rm RX})
\label{eq:rx_phase_socp_obj}
\\
\text{s.t.}\quad
&
\sqrt{\eta_{\rm RX}}\,
\|\mathbf z_{\rm RX}(\mathbf u_{\rm RX})\|_2
\le 1,
\label{eq:rx_phase_socp_si}
\\
&
0\le u_j^{\rm RX}\le 1,
\quad j=1,\ldots,N_{\rm RX},
\label{eq:rx_phase_socp_box}
\\
&
\sum_{j=1}^{N_{\rm RX}}u_j^{\rm RX}
\ge 1.
\label{eq:rx_phase_socp_min}
\end{align}
\end{subequations}
For fixed $\phi_q$ and $\mathbf a_{\rm TX}$, this is a convex \ac{SOCP}, by
the same argument as for the \ac{TX} subproblem.

% ======================================================
\subsubsection{Deterministic Rounding and Alternating Procedure}
\label{subsec:phase_socp_rounding}
% ======================================================

After solving a \ac{TX} or \ac{RX} block problem, the relaxed activation vector
is converted into a finite set of binary trial patterns. Given
$\mathbf u\in[0,1]^N$, let $\pi$ be a permutation such that
$u_{\pi_1}\ge u_{\pi_2}\ge\cdots\ge u_{\pi_N}$. For each cardinality
$k=1,\ldots,N$, define the top-$k$ rounded vector
\begin{equation}
[\mathcal R_k(\mathbf u)]_{\pi_m}
=
\begin{cases}
1, & m\le k,\\
0, & m>k.
\end{cases}
\label{eq:topk_rounding}
\end{equation}
We also form the threshold-rounded vector
\begin{equation}
[\mathcal R_{\rm th}(\mathbf u)]_i
=
\mathbf{1}_{\{u_i\ge 0.5\}},
\label{eq:threshold_rounding}
\end{equation}
and, if it is empty, activate its largest entry. Duplicate trial patterns are
removed.

The complete alternating procedure is summarized in
Algorithm~\ref{alg:phase_quant_socp}. We use deterministic rounding so that
each relaxed solution yields a reproducible, finite set of trial activation
patterns with different numbers of active \acp{PE}. Each trial pattern is then
evaluated under the practical model. A block update is accepted only if it
satisfies the constraint in \eqref{eq:practical_hard_problem} and strictly improves the sum spectral efficiency \eqref{eq:sum_rate_original}.

\IncMargin{0.55em}
\begin{algorithm}[t!]
\caption{Phase-Anchored \acs{SOCP} with Deterministic Rounding}
\label{alg:phase_quant_socp}
\KwIn{Initial activation $(\mathbf a_{\rm TX},\mathbf a_{\rm RX})$ feasible under the practical model, phase-grid size $Q_\phi$, guard margin $\Delta_{\rm guard}$}
\KwOut{Activation $(\mathbf a_{\rm TX},\mathbf a_{\rm RX})$ feasible under the practical model}

Set $P_{\rm SI,max}^{\rm sm}
=P_{\rm SI,max}^{\rm pr}10^{-\Delta_{\rm guard}/10}$\;
Set improvement tolerance $\epsilon_{\rm imp}=10^{-12}$\;

\Repeat{no accepted block update improves the sum spectral efficiency}{
    \ForEach{$s\in\{\mathrm{TX},\mathrm{RX}\}$}{
        Fix the opposite-side activation and set $\mathcal T_s\leftarrow\emptyset$\;

        \For{$q=0,\ldots,Q_\phi-1$}{
            Solve the corresponding phase-anchored \acs{SOCP} block problem\;
            Let $\mathbf u_s^{(q)}$ be the relaxed solution\;
            Add $\mathcal R_{\rm th}(\mathbf u_s^{(q)})$ and all
            $\mathcal R_k(\mathbf u_s^{(q)})$ to $\mathcal T_s$\;
        }

        Remove duplicate trial patterns from $\mathcal T_s$\;
        Discard trial patterns violating
        $P_{\rm TX}|H_{\rm SI}|^2\le P_{\rm SI,max}^{\rm pr}$
        under the practical model\;
        Accept the feasible trial pattern with the largest sum spectral efficiency only if it improves the incumbent by more than $\epsilon_{\rm imp}$\;
    }
}

\end{algorithm}
\DecMargin{0.55em}

For the default setting $N_{\rm TX}=N_{\rm RX}=13$, the deterministic rounding
enumerates the top-$k$ binary activations for $k=1,\ldots,13$. The
threshold-rounded activation is included in this set after duplicate removal.
Hence, each phase anchor produces at most $13$ distinct trial patterns per
\ac{TX} or \ac{RX} block. With $Q_\phi=16$, a block update evaluates at most
$16\times13=208$ rounded trial patterns.

% ======================================================
\subsection{Stage 2: Practical-Model ABLS Refinement}
\label{subsec:practical_considerations}
% ======================================================

Stage~2 applies \ac{ABLS} to the practical model, alternating between the \ac{TX} and \ac{RX} waveguides. On each block, it evaluates all nonempty add, remove, and swap moves relative to the current activation: an add move
activates one currently inactive candidate location, a remove move
deactivates one currently active location, and a swap move simultaneously
deactivates one active location and activates one inactive location. It discards infeasible moves that violate the \ac{SI}-leakage constraint in \eqref{eq:practical_hard_problem}, and accepts only the move that yields the largest improvement in the sum spectral efficiency \eqref{eq:sum_rate_original}. In the proposed method, \ac{ABLS} refines the activation returned by Stage~1; as a standalone baseline, it is applied directly to the common seed that is feasible under the practical model. The procedure is summarized in Algorithm~\ref{alg:real_model_refinement}.

\IncMargin{0.55em}
\begin{algorithm}[t!]
\caption{Practical-Model ABLS Refinement}
\label{alg:real_model_refinement}
\KwIn{Initial activation $(\mathbf a_{\rm TX}^{0},\mathbf a_{\rm RX}^{0})$ feasible under the practical model}
\KwOut{Refined activation $(\mathbf a_{\rm TX},\mathbf a_{\rm RX})$ feasible under the practical model}

Set $(\mathbf a_{\rm TX},\mathbf a_{\rm RX})
\leftarrow(\mathbf a_{\rm TX}^{0},\mathbf a_{\rm RX}^{0})$\;

\Repeat{no accepted sweep improves the sum spectral efficiency}{
    \ForEach{$s\in\{\mathrm{TX},\mathrm{RX}\}$}{
        Fix the opposite-side activation\;
        \For{$r=1,\ldots,I_{\rm local}$}{
            Generate all nonempty add, remove, and swap candidate moves for block $s$\;
            Discard candidate moves violating
            $P_{\rm TX}|H_{\rm SI}|^2\le P_{\rm SI,max}^{\rm pr}$\;
            Evaluate the sum spectral efficiency for the
            remaining candidate moves\;
            Accept the best candidate move only if it improves the incumbent;
            otherwise break\;
        }
    }
}
\end{algorithm}
\DecMargin{0.55em}

Here, $I_{\rm local}$ limits the number of accepted moves within a single
waveguide update, not the total number over the entire procedure. A \ac{TX} or
\ac{RX} update terminates either after $I_{\rm local}$ accepted moves or as soon
as no feasible neighboring activation strictly improves the sum spectral
efficiency \eqref{eq:sum_rate_original}. One complete sweep consists of a
\ac{TX} update followed by an \ac{RX} update, and sweeps are repeated until
neither waveguide is updated during a complete sweep.

The proposed method and the standalone \ac{ABLS} baseline use the same
practical-model neighborhood, \ac{SI}-leakage feasibility test,
best-improvement acceptance rule, and stopping criterion. They differ only in
their initial activations: the proposed method starts from the activation
returned by Stage~1, whereas standalone \ac{ABLS} starts from the common seed,
which is feasible under the practical model. The difference in their final sum
spectral efficiencies is therefore attributable to the Stage~1 \ac{SOCP}
initialization.

% ======================================================
\subsection{Computational Complexity}
\label{subsec:complexity_summary}
% ======================================================

We characterize the dominant per-run computational costs of the proposed
method and contrast them with exhaustive search. The analysis assumes that
each candidate activation is evaluated under the practical model.
Such an evaluation requires forming the effective downlink, uplink, and
cross-waveguide \ac{SI} channels. The dominant operation is the bilinear form
$\mathbf y_{\rm RX}^{T}\mathbf G_{\rm SI}\mathbf y_{\rm TX}$, and hence
\begin{equation}
C_{\rm eval}
=
\mathcal O\!\left(
N_{\rm TX}N_{\rm RX}
+N_{\rm TX}+N_{\rm RX}
\right)
=
\mathcal O\!\left(N_{\rm TX}N_{\rm RX}\right).
\label{eq:eval_complexity}
\end{equation}

\emph{Exhaustive search.}
With at least one active \ac{PE} required on each waveguide, exhaustive search
evaluates
$(2^{N_{\rm TX}}-1)(2^{N_{\rm RX}}-1)$ admissible joint activations.
Its computational cost is therefore
\begin{equation}
\begin{aligned}
C_{\rm exh}
&=
\mathcal O\!\left(
(2^{N_{\rm TX}}-1)(2^{N_{\rm RX}}-1)C_{\rm eval}
\right)
\\
&=
\mathcal O\!\left(
2^{N_{\rm TX}+N_{\rm RX}}
N_{\rm TX}N_{\rm RX}
\right).
\end{aligned}
\label{eq:exhaustive_complexity}
\end{equation}
This exponential growth restricts exhaustive search to moderate candidate
grids.

\emph{Stage~1: phase-anchored \ac{SOCP} initialization.}
A generic \ac{SOCP} with $N$ optimization variables can be solved by an
interior-point method with computational complexity
$\mathcal O(N^{3.5})$ \cite{lobo1998socp,boyd2004convex}. For each waveguide
and each phase anchor, Stage~1 solves one such \ac{SOCP} and evaluates at most
$N$ distinct rounded trial activations under the practical model.
Consequently, over $I_{\rm S1}$ alternating sweeps and $Q_\phi$ phase anchors,
the dominant Stage~1 cost is
\begin{equation}
\begin{aligned}
C_{\rm S1}
=
\mathcal O\Big(
I_{\rm S1}Q_\phi\big[
&N_{\rm TX}^{3.5}
+N_{\rm RX}^{3.5}
\\
&+
(N_{\rm TX}+N_{\rm RX})N_{\rm TX}N_{\rm RX}
\big]
\Big).
\end{aligned}
\label{eq:phase_socp_total_complexity}
\end{equation}
Here the interior-point \ac{SOCP} solve dominates each per-anchor cost, since
$N^{3.5}$ exceeds the $\mathcal O(N^3)$ trial-screening term on a square grid. The sorting operations required by deterministic rounding have complexity
$\mathcal O(N_{\rm TX}\log N_{\rm TX}
+N_{\rm RX}\log N_{\rm RX})$ per phase anchor and are lower order than
the practical-model trial evaluations.

\emph{Stage~2: practical-model \ac{ABLS} refinement.}
For a waveguide with $N$ candidate locations, the neighborhood is dominated by
the $\mathcal O(N^2)$ swap moves, so one best-improvement iteration requires at
most $\mathcal O(N^2)$ practical-model evaluations. Since each \ac{TX} and \ac{RX} block update accepts at most
$I_{\rm local}$ moves, the dominant cost over $I_{\rm S2}$ complete
refinement sweeps is
\begin{equation}
C_{\rm S2}
=
\mathcal O\!\left(
I_{\rm S2}I_{\rm local}
(N_{\rm TX}^{2}+N_{\rm RX}^{2})
N_{\rm TX}N_{\rm RX}
\right).
\label{eq:refinement_complexity}
\end{equation}

Per sweep, both stages are polynomial in the candidate-grid dimensions. The
sweep counts $I_{\rm S1}$ and $I_{\rm S2}$ are not bounded a priori, but every
accepted update strictly increases the objective over the finite binary
feasible set, which guarantees finite termination: Stage~1 stops when no
rounded trial activation improves the incumbent, and Stage~2 stops at a
blockwise local optimum over the add, remove, and swap neighborhoods. In the
experiments of Section~\ref{subsec:candidate_density_scalability}, both counts
remain small, and the measured runtime scales polynomially up to $N=60$.

The proposed method thus replaces the exponential exhaustive-search cost
\eqref{eq:exhaustive_complexity} with per-sweep costs polynomial in the grid
size, which is what enables evaluation up to $N=60$ in
Section~\ref{subsec:candidate_density_scalability}, where exhaustive search is
infeasible. Table~\ref{tab:complexity} summarizes the operation counts for a
square candidate grid with $N_{\rm TX}=N_{\rm RX}=N$.

\begin{table}[t]
\centering
\caption{Parameterized per-run computational costs for a square candidate grid
$N=N_{\rm TX}=N_{\rm RX}$.}
\label{tab:complexity}
\renewcommand{\arraystretch}{1.3}
\setlength{\tabcolsep}{3pt}
\begin{tabular}{@{}p{0.30\columnwidth}p{0.62\columnwidth}@{}}
\hline
\textbf{Method} & \textbf{Dominant cost} \\
\hline
Exhaustive search
&
$\mathcal O\!\left(4^N N^2\right)$
\\[0.25em]
Proposed, Stage~1
&
$\mathcal O\!\left(
I_{\rm S1}Q_\phi
\left[N^{3.5}+N^3\right]
\right)$
\\[0.25em]
Proposed, Stage~2
&
$\mathcal O\!\left(
I_{\rm S2}I_{\rm local}N^4
\right)$
\\
\hline
\end{tabular}
\end{table}

% ======================================================
\section{Numerical Results}
\label{sec:numerical_results}
% ======================================================

This section evaluates the proposed phase-anchored \ac{SOCP}+\ac{ABLS} method.
Its \ac{SOCP} initialization uses the simplified model, whereas its \ac{ABLS}
refinement runs under the practical model. All reported spectral efficiencies
are computed under the practical FullPASS model and only for activations
feasible in \eqref{eq:practical_hard_problem}, so FullPASS-side \ac{SI} leakage
is an admissibility condition rather than an additive rate penalty.

% ======================================================
\subsection{Simulation Setup}
\label{subsec:simulation_setup}
% ======================================================

The nominal simulation parameters are summarized in
Table~\ref{tab:simulation_parameters}. The full-duplex user-terminal coordinate follows
\eqref{eq:fd_user_coordinate}, with
$x_u\in[0,1.2\max\{L_{\rm TX},L_{\rm RX}\}]$ and
$y_u\in[-4D,4D]$. The main comparison averages over $150$ random
full-duplex user-terminal scenarios. For each parameter sweep and for the
large-array runtime study, this same fixed bank of scenarios is reused at
every sweep point, so that changes in the reported curves are caused by the
varied system parameter rather than by different user realizations. Reported runtimes include only
method execution; common channel-table construction and empirical guard
calibration are treated as offline preprocessing. The continuous \ac{SOCP} relaxations are implemented in CVXPY~\cite{diamond2016cvxpy} and solved using
Clarabel~\cite{goulart2024clarabel}.

All reported performance uses the desired-link sum spectral efficiency in \eqref{eq:sum_rate_original}, evaluated under the practical FullPASS model.

For each final activation that is feasible under the practical model, the achieved desired-link sum spectral efficiency is
\begin{align}
R_{\rm sum}
=
&\log_2\left(1+\frac{P_{\rm DL}}{5.00\times10^{-14}}\right)
\nonumber\\[-0.15em]
+&\log_2\left(1+\frac{P_{\rm UL}}{5.00\times10^{-14}}\right)
\quad \mathrm{bit/s/Hz}.
\label{eq:spectral_efficiency_numerical}
\end{align}

The nominal parameters in Table~\ref{tab:simulation_parameters} are used as a
representative room-scale FullPASS deployment. 

\begin{table}[t!]
\centering
\caption{Nominal simulation parameters.}
\label{tab:simulation_parameters}
\renewcommand{\arraystretch}{1.08}
\begin{tabular}{@{}p{0.50\columnwidth}p{0.38\columnwidth}@{}}
\hline
Parameter & Value \\
\hline
Carrier frequency $f_0$ & $14~\mathrm{GHz}$ \\
Wavelength $\lambda_0$ & $c/f_0$ \\
Effective refractive index $n_e$ & $1.6$ \\
Waveguide lengths $L_{\rm TX},L_{\rm RX}$ & $20,20~\mathrm{m}$ \\
Waveguide separation $D$ & $1~\mathrm{m}$ \\
In-waveguide attenuation $\alpha_{\rm TX},\alpha_{\rm RX}$ & $1.2,1.2~\mathrm{dB/m}$ \\
Candidate locations $N_{\rm TX},N_{\rm RX}$ & $13,13$ \\
FullPASS transmit power $P_{\rm TX}$ & $30~\mathrm{dBm}$ \\
Full-duplex user-terminal transmit power $P_{\rm U}$ & $20~\mathrm{dBm}$ \\
Receiver noise figure $\mathrm{NF}$ & $4~\mathrm{dB}$ \\
Effective noise-plus-interference powers
$\sigma_{\rm DL}^2,\sigma_{\rm UL}^2$ &
$-103.01~\mathrm{dBm}$ ($5.00\!\times\!10^{-14}~\mathrm{W}$) \\
Practical FullPASS-side \ac{SI} limit $P_{\rm SI,max}^{\rm pr}$ & $-60~\mathrm{dBm}$ \\
Pass-through coefficients $\tau_{\rm TX},\tau_{\rm RX}$ & $0.97,0.97$ \\
Coupling coefficients $\delta_{\rm TX},\delta_{\rm RX}$ &
$\sqrt{1-\tau_{\rm TX}^2},\sqrt{1-\tau_{\rm RX}^2}$ \\
User-plane height $H$ & $2.4~\mathrm{m}$ \\
Reported desired-link metric & $R_{\rm sum}$ in $\mathrm{bit/s/Hz}$ \\
Practical \ac{SI} treatment & Leakage constraint \\
Nonempty-side condition & At least one \ac{PE} per waveguide \\
Phase-grid size $Q_\phi$ & $16$ \\
Guard mismatch target $\epsilon$ & $0.05$ \\
Guard-calibration samples & $50{,}000$ \\
Maximum accepted \ac{ABLS} moves per block call $I_{\rm local}$ & $20$ \\
\ac{SOCP} solver & Clarabel \\
Number of user scenarios & $150$ \\
\hline
\end{tabular}
\end{table}

\subsection{Compared Methods}

Three methods are considered. \emph{Exhaustive search} evaluates the
finite-grid optimum of \eqref{eq:practical_hard_problem} and serves as the
benchmark on candidate grids for which full enumeration is computationally feasible. \emph{Phase-anchored \ac{SOCP}+\ac{ABLS}} is
the proposed method. Finally, \emph{standalone \ac{ABLS}} is a direct practical-model
local-search baseline. All iterative methods start from the same activation that is feasible under the practical model.

For simplified-model trial generation, the empirical guard calibration gives
$\Delta_{\rm guard}=0.43~\mathrm{dB}$ for the target mismatch probability
$\epsilon=0.05$.

% ======================================================
\subsection{Moderate-Grid Near-Optimality and Runtime Comparison}
\label{subsec:runtime_quality_tradeoff}
% ======================================================

\begin{figure*}[t!]
\centering
\begin{minipage}[t]{0.48\textwidth}
\centering
\includegraphics[
    width=\columnwidth,
    height=0.16\textheight
]{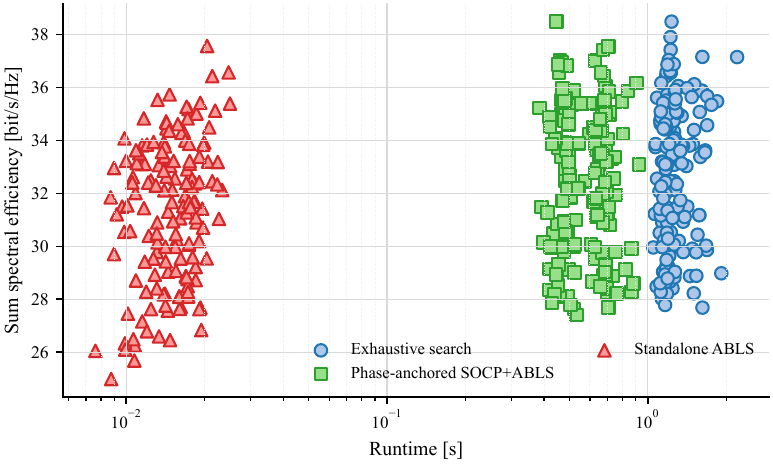}

\vspace{0.2em}
\small (a) $N_{\rm TX}=N_{\rm RX}=13$
\end{minipage}\hfill
\begin{minipage}[t]{0.48\textwidth}
\centering
\includegraphics[
    width=\columnwidth,
    height=0.16\textheight
]{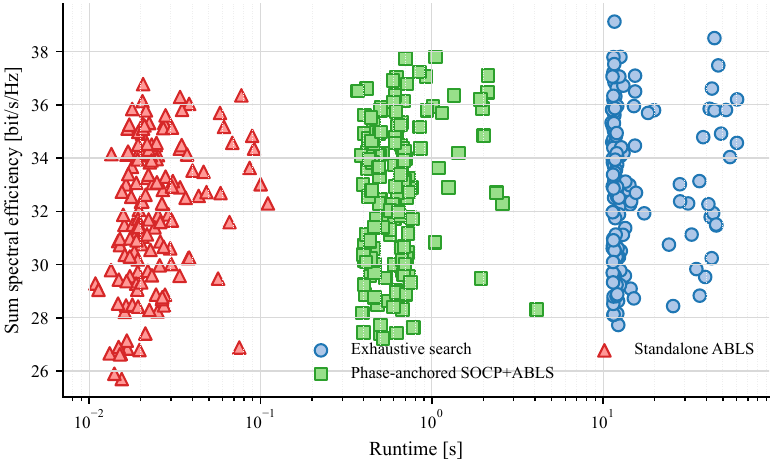}

\vspace{0.2em}
\small (b) $N_{\rm TX}=N_{\rm RX}=15$
\end{minipage}
\caption{Scenario-wise runtime--spectral-efficiency comparison. Exhaustive search is included as a finite-grid
benchmark.}
\label{fig:moderate_grid_tradeoff}
\end{figure*}

Fig.~\ref{fig:moderate_grid_tradeoff}(a) and (b) compare the three methods on the
nominal $13\times13$ and $15\times15$ candidate grids, i.e.,
$N_{\rm TX}=N_{\rm RX}=13$ and $N_{\rm TX}=N_{\rm RX}=15$.

On the $13\times13$ grid, exhaustive search provides the finite-grid optimum.
It achieves an average desired-link sum spectral efficiency of
$32.705~\mathrm{bit/s/Hz}$ with an average runtime of $1.233~\mathrm{s}$.
The proposed phase-anchored \ac{SOCP}+\ac{ABLS} method achieves
$32.395~\mathrm{bit/s/Hz}$, corresponding to a $0.95\%$ gap to exhaustive
search, while reducing the average runtime by a factor of $2.29$. In contrast,
standalone \ac{ABLS} has the lowest average runtime of $0.016~\mathrm{s}$ but
achieves $31.462~\mathrm{bit/s/Hz}$, with a $3.80\%$ gap to exhaustive search.
Thus, relative to standalone \ac{ABLS}, the proposed initialization improves
the average spectral efficiency by $0.933~\mathrm{bit/s/Hz}$ and recovers
approximately $75\%$ of the gap between standalone \ac{ABLS} and exhaustive
search.

The $15\times15$ comparison further assesses the regime in which exhaustive
search becomes increasingly expensive. Increasing the grid size from
$13\times13$ to $15\times15$ increases the number of nonempty joint activation
patterns from $(2^{13}-1)^2$ to $(2^{15}-1)^2$. Nevertheless, the proposed
phase-anchored \ac{SOCP}+\ac{ABLS} method remains within $0.95\%$ of the
exhaustive-search spectral efficiency at this larger grid. Exhaustive search
requires $11.08~\mathrm{s}$ on average, whereas the proposed method requires
more than $10\times$ lower average runtime. Thus, the proposed initialization
retains near-optimal desired-link performance while avoiding the rapidly
increasing cost of exhaustive enumeration.

% ======================================================
\subsection{Candidate Density and Scalability}
\label{subsec:candidate_density_scalability}
% ======================================================

Exhaustive search does not extend beyond $N=15$ because its search space
grows exponentially with the number of candidate locations. In particular,
for $N_{\rm TX}=N_{\rm RX}=N$, it must examine $(2^N-1)^2$ nonempty joint
activation patterns; this number is already approximately $1.07\times10^9$ at
$N=15$ and increases to approximately $1.15\times10^{18}$ at $N=30$.
Section~\ref{subsec:runtime_quality_tradeoff} already validated the solution
quality of the proposed method against this exhaustive benchmark at
$N=13$ and $15$; here we extend the evaluation to substantially larger
candidate densities, where exhaustive search is no longer computationally
evaluated. Therefore, exhaustive search is used only as a finite-grid
benchmark at small and moderate candidate densities.

In contrast, the proposed phase-anchored \ac{SOCP}+\ac{ABLS} method does not
enumerate the full joint activation space (Section~\ref{subsec:complexity_summary}),
so its cost grows far more slowly. Thus, similar runtimes on small grids do not
imply similar computational scaling: the fixed \ac{SOCP} overhead is visible
at small $N$, whereas exhaustive-search cost dominates at larger $N$.

\begin{figure*}[t!]
\centering
\begin{minipage}[t]{0.48\textwidth}
\centering
\includegraphics[
    width=\columnwidth,
    height=0.16\textheight
]{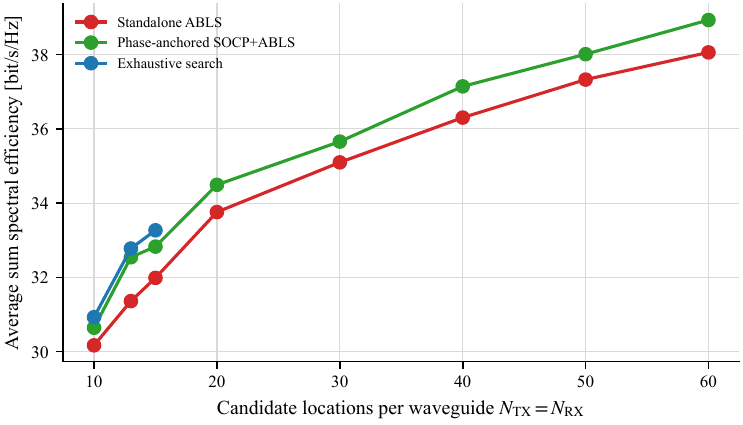}

\vspace{0.2em}
\small (a) Desired-link sum spectral efficiency
\end{minipage}\hfill
\begin{minipage}[t]{0.48\textwidth}
\centering
\includegraphics[
    width=\columnwidth,
    height=0.16\textheight
]{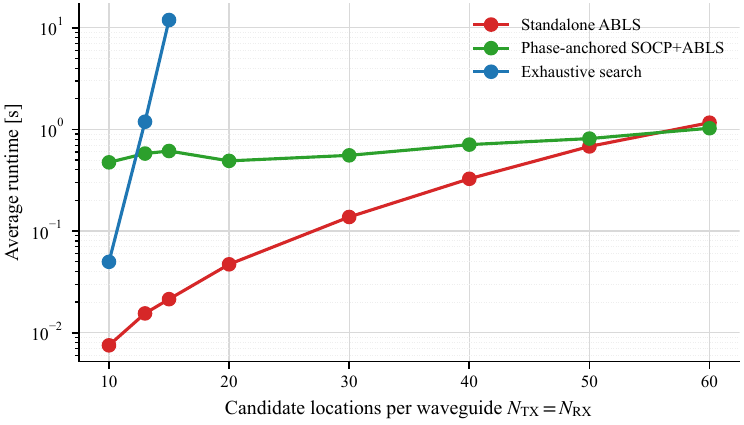}

\vspace{0.2em}
\small (b) Average runtime
\end{minipage}
\caption{Candidate-density and scalability evaluation under the practical-model
\ac{SI}-leakage constraint. Left: achieved desired-link sum spectral
efficiency versus the number of candidate locations per waveguide. Right:
average runtime. Exhaustive search is evaluated only through $N=15$, the largest grid size for which exhaustive search was evaluated in our
simulations.}
\label{fig:scalability_summary}
\end{figure*}

Fig.~\ref{fig:scalability_summary}(a) shows that increasing the candidate
density improves the achievable desired-link spectral efficiency by providing
finer geometry control along waveguides of fixed length, and that both scalable
methods benefit from the additional candidate locations.

Fig.~\ref{fig:scalability_summary}(b) shows the corresponding average runtime.
Exhaustive-search runtime increases rapidly from approximately
$0.03~\mathrm{s}$ at $N=10$ to $11.08~\mathrm{s}$ at $N=15$ and is therefore
not evaluated for larger grids. In contrast, the proposed method remains
applicable up to $N=60$ and stays below $1~\mathrm{s}$ on average at the
largest tested grid. In this larger-grid regime where complete enumeration is
impractical, it retains its desired-link advantage over standalone \ac{ABLS}.

% ======================================================
\subsection{Full-Duplex Gain over Same-Hardware Half Duplex}
\label{subsec:fd_vs_hd_candidate_density}
% ======================================================

We next compare FullPASS with a same-hardware \ac{HD}-\ac{TDD} reference. Both systems use the same two waveguides,
candidate locations, practical propagation model,
in-waveguide attenuation, \ac{PE} coupling coefficients, and $150$ common
user-terminal scenarios. In each time slot, the transmitting side uses the
same power as in the full-duplex case: $P_{\rm TX}$ during the downlink slot
and $P_{\rm U}$ during the uplink slot. For each candidate density, the
\ac{HD}-\ac{TDD} reference independently selects the downlink transmit-side
and uplink receive-side activation patterns by exhaustive search under the
practical model. Since the two directions occupy separate time intervals, this
reference has no simultaneous FullPASS-side transmit--receive coupling and
therefore no simultaneous \ac{SI}-leakage constraint.

Equal downlink and uplink time fractions are used to represent bidirectional
service. Hence, the time-averaged desired-link spectral efficiency of the
half-duplex reference is
\begin{equation}
R_{\rm HD}
=
\frac{1}{2}
\log_2\left(
1+\frac{P_{\rm DL}^{\rm HD}}{\sigma_{\rm DL}^2}
\right)
+
\frac{1}{2}
\log_2\left(
1+\frac{P_{\rm UL}^{\rm HD}}{\sigma_{\rm UL}^2}
\right).
\label{eq:half_duplex_rate}
\end{equation}
In contrast, the FullPASS methods operate in \ac{IBFD} mode and report
$R_{\rm FD}=R_{\rm DL}+R_{\rm UL}$, while requiring the FullPASS-side \ac{SI} leakage to satisfy the constraint in
\eqref{eq:practical_hard_problem}.

\begin{figure}[t!]
\centering
\includegraphics[
    width=0.95\columnwidth,
    height=0.16\textheight
]
{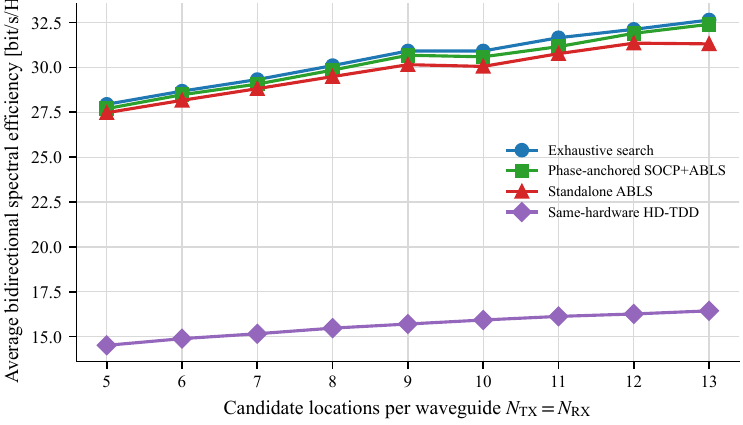}
\caption{Average bidirectional spectral efficiency versus the number of
candidate locations per waveguide. Full-duplex methods jointly select transmit
and receive \acp{PE} under the practical-model \ac{SI}-leakage constraint.
The same-hardware \ac{HD}-\ac{TDD} reference independently optimizes the
downlink and uplink selections by exhaustive search and uses equal downlink
and uplink time fractions.}
\label{fig:fd_hd_candidate_density}
\end{figure}

Fig.~\ref{fig:fd_hd_candidate_density} compares the resulting average
bidirectional spectral efficiencies for
$N_{\rm TX}=N_{\rm RX}=5,\ldots,13$. The \ac{HD}-\ac{TDD} reference
increases from $14.51~\mathrm{bit/s/Hz}$ at $N=5$ to
$16.44~\mathrm{bit/s/Hz}$ at $N=13$. Over the same range, the proposed
phase-anchored \ac{SOCP}+\ac{ABLS} method increases from
$27.71~\mathrm{bit/s/Hz}$ to $32.41~\mathrm{bit/s/Hz}$. Consequently, it
provides a $90.9\%$--$97.2\%$ increase in average bidirectional spectral
efficiency relative to the same-hardware half-duplex reference.

The full-duplex exhaustive-search benchmark provides a
$92.6\%$--$98.6\%$ gain over the half-duplex reference, whereas standalone
\ac{ABLS} provides an $88.8\%$--$92.8\%$ gain. The proposed method remains
within $1.57\%$ of the full-duplex exhaustive-search benchmark across the
examined candidate densities and consistently improves on standalone \ac{ABLS}.
These results show that the gain of FullPASS is not merely due to having two
waveguides: even against a half-duplex system with the same physical hardware,
independent per-direction exhaustive activation selection, and no simultaneous
\ac{SI}, geometry-based passive \ac{SI} suppression enables the two directions
to reuse the same time--frequency resource and yields a substantial
bidirectional spectral efficiency advantage.

% ======================================================
\subsection{Impact of the \ac{SI}-Suppression Requirement}
\label{subsec:si_requirement_sensitivity}
% ======================================================

We next examine how the FullPASS-side \ac{SI}-suppression requirement
affects the achievable desired-link performance. We define the suppression
requirement as
\begin{equation}
\Gamma_{\rm SI}
=
P_{\rm TX}^{\rm dBm}-P_{\rm SI,max}^{\rm pr,dBm}.
\label{eq:si_suppression_requirement_sweep}
\end{equation}
A larger $\Gamma_{\rm SI}$ corresponds to a lower permitted \ac{SI}
leakage power and therefore a stricter \ac{SI}-leakage constraint. For
the nominal transmit power $P_{\rm TX}=30~\mathrm{dBm}$, the maximum allowable \ac{SI}-leakage power $P_{\rm SI,max}^{\rm pr}=-60~\mathrm{dBm}$ corresponds to
$\Gamma_{\rm SI}=90~\mathrm{dB}$.

\begin{figure}[t!]
\centering
\includegraphics[
    width=0.95\columnwidth,
    height=0.16\textheight
]
{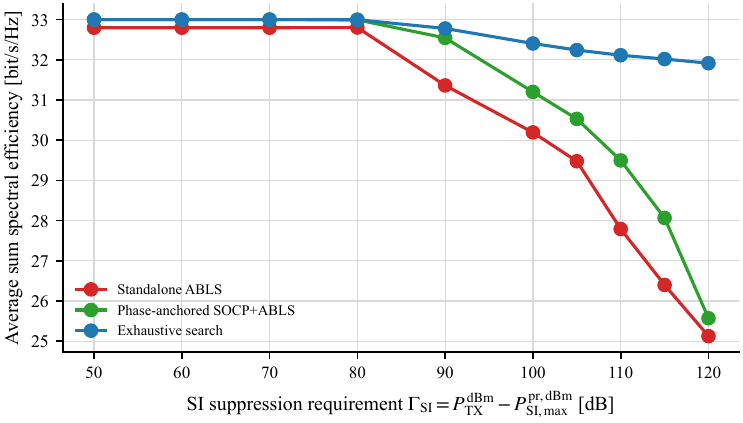}
\caption{Average desired-link sum spectral efficiency versus the FullPASS-side \ac{SI}-suppression requirement $\Gamma_{\rm SI}$. Larger
$\Gamma_{\rm SI}$ corresponds to a stricter practical-model
\ac{SI}-leakage constraint.}
\label{fig:si_requirement_tradeoff}
\end{figure}

Fig.~\ref{fig:si_requirement_tradeoff} shows that the \ac{SI}
suppression requirement has little effect when it is loose. In this regime, the
high-quality activation patterns selected for the desired uplink and downlink
already satisfy the \ac{SI} constraint. Consequently, the hard
constraint does not substantially restrict the available geometry choices.

As $\Gamma_{\rm SI}$ increases, the \ac{SI} power constraint becomes
restrictive. The activation pattern must then reduce the coherent coupling
between the transmit and receive waveguides, which limits the freedom to
independently maximize the desired downlink and uplink channel gains. The
resulting reduction in desired-link spectral efficiency illustrates the
fundamental tradeoff in FullPASS: stronger \ac{SI} suppression requirement leaves
fewer favorable transmit--receive pinching-element combinations.

The proposed phase-anchored \ac{SOCP}+\ac{ABLS} method remains closer to the
exhaustive-search benchmark than standalone \ac{ABLS} as the requirement
becomes more stringent. This indicates that the phase-anchored initialization
helps identify higher-quality activation patterns that satisfy the practical-model \ac{SI}-leakage constraint when that constraint substantially restricts the discrete search space.

% ======================================================
\subsection{Impact of \ac{PE} Coupling Ratio}
\label{subsec:coupling_sensitivity}
% ======================================================

The per-\ac{PE} power-coupling ratio is defined as
\begin{equation}
\kappa=1-\tau^2,
\label{eq:kappa_coupling_ratio}
\end{equation}
where $\delta=\sqrt{\kappa}$ and $\tau=\sqrt{1-\kappa}$. Increasing $\kappa$ increases the fraction of guided power coupled between the
waveguide and free space at each active \ac{PE}. At the same time, it reduces
the guided power available for downstream \acp{PE} and increases the aggregate
transmit-to-receive coupling.

\begin{figure}[t!]
\centering
\includegraphics[
    width=0.95\columnwidth,
    height=0.16\textheight
]
{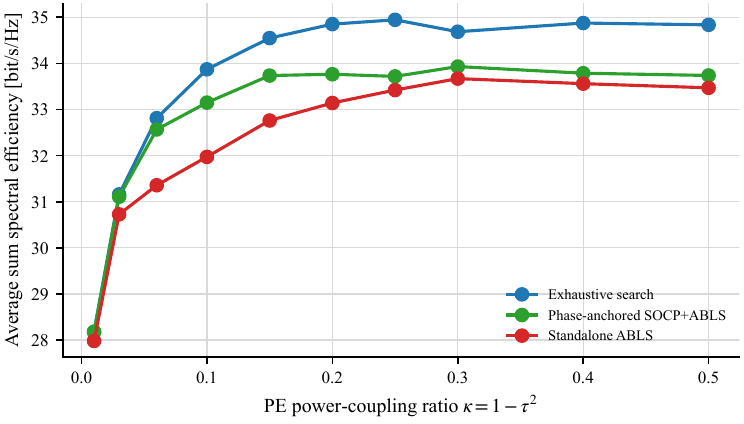}
\caption{Average desired-link sum spectral efficiency versus the per-\ac{PE}
power-coupling ratio $\kappa=1-\tau^2$.}
\label{fig:pa_coupling_tradeoff}
\end{figure}

Fig.~\ref{fig:pa_coupling_tradeoff} shows that the desired-link spectral
efficiency increases rapidly at low coupling ratios. In this regime, the desired links are mainly limited by the small fraction of
guided power coupled at each active \ac{PE}. As $\kappa$ increases further, the spectral-efficiency gain becomes
small and eventually saturates. This behavior reflects the competing effects
of stronger local coupling, reduced pass-through to downstream \acp{PE}, and
the more restrictive \ac{SI}-leakage constraint.

The advantage of the phase-anchored \ac{SOCP} initialization narrows as the
per-\ac{PE} coupling strength increases. A larger $\kappa$ corresponds to a
smaller pass-through coefficient $\tau$, making the practical channel gains
more dependent on the ordered pattern of upstream activated \acp{PE}. Since
the \ac{SOCP} initialization uses the simplified model without this
activation-dependent cumulative pass-through loss, its trial activation patterns become less
representative of the practical optimization problem at strong coupling.
Consequently, the subsequent \ac{ABLS} refinement has less benefit from the
\ac{SOCP}-generated initialization in this regime.

% ======================================================
\subsection{Impact of In-Waveguide Attenuation}
\label{subsec:waveguide_loss_diagnostic}
% ======================================================

We next evaluate the effect of the continuous guided-wave attenuation
coefficient $\alpha=\alpha_{\rm TX}=\alpha_{\rm RX}$ while keeping the
remaining nominal parameters fixed. A larger $\alpha$ reduces the guided power
available at remote \acp{PE}, thereby changing both the desired-link gains and
the preferred activation geometry.

\begin{figure}[t!]
\centering
\includegraphics[
    width=0.95\columnwidth,
    height=0.16\textheight
]
{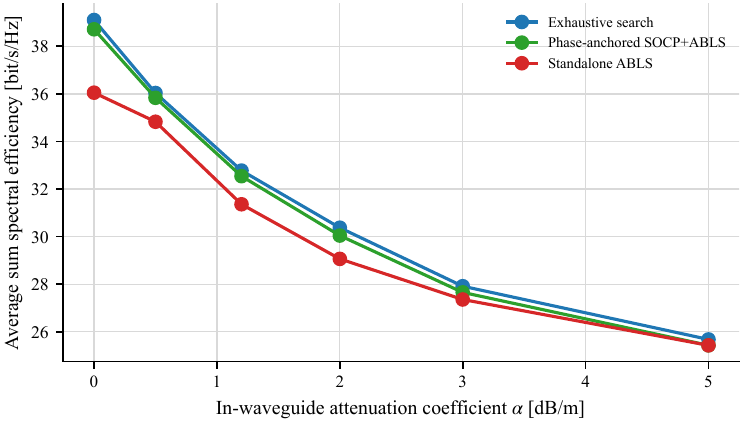}
\caption{Average desired-link sum spectral efficiency versus the guided-wave
attenuation coefficient
$\alpha=\alpha_{\rm TX}=\alpha_{\rm RX}$.}
\label{fig:waveguide_loss_sensitivity}
\end{figure}

Fig.~\ref{fig:waveguide_loss_sensitivity} shows that increasing $\alpha$
reduces the usefulness of remote \acp{PE} and therefore lowers the available
geometry gain. The proposed method decreases from
$38.78~\mathrm{bit/s/Hz}$ at $\alpha=0~\mathrm{dB/m}$ to
$25.44~\mathrm{bit/s/Hz}$ at $\alpha=5~\mathrm{dB/m}$.

Beyond lowering the link budget, in-waveguide attenuation also reduces the
usefulness of distributed geometry control along the waveguides.

% ======================================================
\section{Conclusion}
\label{sec:conclusion}
% ======================================================

This paper introduced FullPASS, a dual-waveguide pinching-antenna architecture
for in-band full-duplex communication with a single full-duplex user terminal,
in which the selected transmit and receive activations jointly shape the desired
downlink and uplink channels and the cross-waveguide \ac{SI} coupling. Using a
practical geometry-based propagation model, we formulated joint \ac{PE} selection as a
nonconvex binary problem that maximizes the bidirectional sum spectral
efficiency subject to an \ac{SI}-leakage constraint, and solved it with a
two-stage method combining a phase-anchored \ac{SOCP} relaxation and
deterministic rounding with practical-model \ac{ABLS} refinement.

Because exhaustive search attains the finite-grid optimum but scales exponentially and is therefore confined to small
candidate grids, the proposed method replaces this enumeration with a finite set
of phase-anchored \ac{SOCP} subproblems and low-order local moves, at almost no
cost in quality: it attains a desired-link sum spectral efficiency within
$0.95\%$ of exhaustive search on the $13\times13$ and $15\times15$ grids, at more
than an order of magnitude lower runtime on the latter, and remains applicable up
to $N_{\rm TX}=N_{\rm RX}=60$, beyond the grid sizes evaluated for exhaustive
search. The parameter studies expose the central FullPASS tradeoff: stronger
\ac{SI} suppression and larger in-waveguide attenuation each shrink the set of
favorable transmit--receive geometries, while stronger per-\ac{PE} coupling helps
only up to a saturation point, showing that \ac{PE} selection balances
desired-link quality against self-interference.

% ======================================================
\bibliographystyle{IEEEtran}
\bibliography{refs}

\end{document}